\documentclass[12pt]{article}
\usepackage{amsfonts}
\usepackage{amssymb}
\usepackage{graphics,psboxit,amsmath}
\usepackage{subfigure}
\usepackage{graphicx}
\usepackage{verbatim}
\usepackage[Euler]{upgreek}

\def\hybrid{\topmargin -30pt    \oddsidemargin 0pt 
        \headheight 0pt \headsep 0pt
        \textwidth 6.25in       
        \textheight 9.5in       
        \marginparwidth .875in
        \parskip 5pt plus 1pt   \jot = 1.5ex}

\hybrid

\def\baselinestretch{1.2}

\catcode`\@=11

\def\marginnote#1{}
\newcount\hour
\newcount\minute
\newtoks\amorpm
\hour=\time\divide\hour by60
\minute=\time{\multiply\hour by60 \global\advance\minute by-\hour}
\edef\standardtime{{\ifnum\hour<12 \global\amorpm={am}%
        \else\global\amorpm={pm}\advance\hour by-12 \fi
        \ifnum\hour=0 \hour=12 \fi
        \number\hour:\ifnum\minute<10 0\fi\number\minute\the\amorpm}}
\edef\militarytime{\number\hour:\ifnum\minute<10 0\fi\number\minute}

\def\draftlabel#1{{\@bsphack\if@filesw {\let\thepage\relax
   \xdef\@gtempa{\write\@auxout{\string
      \newlabel{#1}{{\@currentlabel}{\thepage}}}}}\@gtempa
   \if@nobreak \ifvmode\nobreak\fi\fi\fi\@esphack}
        \gdef\@eqnlabel{#1}}
\def\@eqnlabel{}
\def\@vacuum{}
\def\draftmarginnote#1{\marginpar{\raggedright\scriptsize\tt#1}}

\def\draft{\oddsidemargin -.5truein
        \def\@oddfoot{\sl preliminary draft \hfil
        \rm\thepage\hfil\sl\today\quad\militarytime}
        \let\@evenfoot\@oddfoot \overfullrule 3pt
        \let\label=\draftlabel
        \let\marginnote=\draftmarginnote
   \def\@eqnnum{(\theequation)\rlap{\kern\marginparsep\tt\@eqnlabel}%
\global\let\@eqnlabel\@vacuum}  }

\def\draft2{
        \def\@oddfoot{\sl preliminary draft \hfil
        \rm\thepage\hfil\sl\today\quad\militarytime}
        \let\@evenfoot\@oddfoot \overfullrule 3pt
        \let\label=\draftlabel
        \let\marginnote=\draftmarginnote
   \def\@eqnnum{(\theequation)\rlap{\kern\marginparsep\tt\@eqnlabel}%
\global\let\@eqnlabel\@vacuum}  }

\def\preprint{\twocolumn\sloppy\flushbottom\parindent 2em
        \leftmargini 2em\leftmarginv .5em\leftmarginvi .5em
        \oddsidemargin -.5in    \evensidemargin -.5in
        \columnsep .4in \footheight 0pt
        \textwidth 10.in        \topmargin  -.4in
        \headheight 12pt \topskip .4in
        \textheight 6.9in \footskip 0pt
        \def\@oddhead{\thepage\hfil\addtocounter{page}{1}\thepage}
        \let\@evenhead\@oddhead \def\@oddfoot{} \def\@evenfoot{} }

\def\numberbysection{\@addtoreset{equation}{section}
        \def\theequation{\thesection.\arabic{equation}}}

\def\underline#1{\relax\ifmmode\@@underline#1\else
        $\@@underline{\hbox{#1}}$\relax\fi}

\def\titlepage{\@restonecolfalse\if@twocolumn\@restonecoltrue\onecolumn
     \else \newpage \fi \thispagestyle{empty}\c@page\z@
        \def\thefootnote{\fnsymbol{footnote}} }

\def\endtitlepage{\if@restonecol\twocolumn \else \newpage \fi
        \def\thefootnote{\arabic{footnote}}
        \setcounter{footnote}{0}}  

\catcode`@=12
\relax

\def\figcap{\section*{Figure Captions\markboth
        {FIGURECAPTIONS}{FIGURECAPTIONS}}\list
        {Figure \arabic{enumi}:\hfill}{\settowidth\labelwidth{Figure
999:}
        \leftmargin\labelwidth
        \advance\leftmargin\labelsep\usecounter{enumi}}}
 \relax
\def\tablecap{\section*{Table Captions\markboth
        {TABLECAPTIONS}{TABLECAPTIONS}}\list
        {Table \arabic{enumi}:\hfill}{\settowidth\labelwidth{Table
999:}
        \leftmargin\labelwidth
        \advance\leftmargin\labelsep\usecounter{enumi}}}
 \relax
\def\reflist{\section*{References\markboth
        {REFLIST}{REFLIST}}\list
        {[\arabic{enumi}]\hfill}{\settowidth\labelwidth{[999]}
        \leftmargin\labelwidth
        \advance\leftmargin\labelsep\usecounter{enumi}}}
 \relax
\makeatletter
\newcounter{pubctr}
\def\publist{\@ifnextchar[{\@publist}{\@@publist}}
\def\@publist[#1]{\list
        {[\arabic{pubctr}]\hfill}{\settowidth\labelwidth{[999]}
        \leftmargin\labelwidth
        \advance\leftmargin\labelsep
        \@nmbrlisttrue\def\@listctr{pubctr}
        \setcounter{pubctr}{#1}\addtocounter{pubctr}{-1}}}
\def\@@publist{\list
        {[\arabic{pubctr}]\hfill}{\settowidth\labelwidth{[999]}
        \leftmargin\labelwidth
        \advance\leftmargin\labelsep
        \@nmbrlisttrue\def\@listctr{pubctr}}}
 \relax
\makeatother

\def\be{\begin{equation}}
\def\ee{\end{equation}}
\def\ba{\begin{eqnarray}}
\def\ea{\end{eqnarray}}
\def\ben{\begin{equation*}}
\def\een{\end{equation*}}
\def\ban{\begin{eqnarray*}}
\def\ean{\end{eqnarray*}}

\def\del{\partial}

\def\k{\kappa}

\def\a{\alpha}

\def\b{\beta}

\def\g{\gamma}
\def\G{\Gamma}
\def\d{\delta}

\def\P{\Pi}

\def\th{\theta}
\def\Th{\Theta}
\def\m{\mu}
\def\n{\nu}

\def\Om{\Omega}
\def\l{\lambda}
\def\L{\Lambda}
\def\s{\sigma}
\def\S{\Sigma}

\def\cG{{\cal G}}
\def\cH{{\cal H}}

\def\cN{{\cal N}}
\def\cO{{\cal O}}

\def\elF{{\bf F}}
\def\elK{{\bf K}}
\def\elPi{{\bf \Pi}}
\def\elE{{\bf E}}
\def\real{{\rm Re} \, }
\def\imag{{\rm Im} \, }

\def\no{\noindent}

\def\qq{\qquad}

\def\IR{\relax{\rm I\kern-.18em R}}

\def \ha {{1\over 2}}

\def \ov {\over}

\def\const{{\rm const.}}

\begin{document}


\renewcommand{\theequation}{\thesection.\arabic{equation}}
\csname @addtoreset\endcsname{equation}{section}

\newcommand{\eqn}[1]{(\ref{#1})}
\begin{titlepage}
\begin{center}

\hfill 0706.2655 [hep-th]\\

\vskip .5in

{\Large \bf Stability of string configurations dual\\ to
quarkonium states in AdS/CFT}

\vskip 0.5in

{\bf Spyros D. Avramis$^{1,2}$},\phantom{x} {\bf Konstadinos
Sfetsos}$^1$\phantom{x} and\phantom{x} {\bf Konstadinos Siampos}$^1$
\vskip 0.1in

${}^1\!$
Department of Engineering Sciences, University of Patras,\\
26110 Patras, Greece\\

\vskip .1in

${}^2\!$
Department of Physics, National Technical University of Athens,\\
15773, Athens, Greece\\

\vskip .15in

{\footnotesize {\tt avramis@mail.cern.ch}, \ \ {\tt sfetsos@upatras.gr},
\ \ {\tt ksiampos@upatras.gr}}\\

\end{center}

\vskip .4in

\centerline{\bf Abstract}

\no We extend our earlier work, regarding the perturbative
stability of string configurations used for computing the
interaction potential of heavy quarks within the gauge/gravity
correspondence, to cover a more general class of gravity duals. We
provide results, mostly based on analytic methods and corroborated
by numerical calculations, which apply to strings in a general
class of backgrounds that encompass boosted, spinning and
marginally-deformed D3-brane backgrounds. For the case of spinning
branes we demonstrate in a few examples that perturbative
stability of strings may require strong conditions complementing
those following by thermodynamic stability of the dual field
theories. For marginally-deformed backgrounds, we find that even
in the conformal case stability requires an upper value for the
imaginary part $\s$ of the deformation parameter, whereas in
regions of the Coulomb branch where there exists linear
confinement we find that there exist stable string configurations
for certain ranges of values of the parameter $\s$. We finally
discuss the case of open strings with fixed endpoints propagating
in Rindler space, which turns out to have an exact
classical-mechanical analog.

\vfill
\no


\end{titlepage}
\vfill
\eject


\tableofcontents

\def\baselinestretch{1.2}
\baselineskip 20 pt
\no

\newpage

\section{Introduction}

The AdS/CFT correspondence \cite{adscft} maps the computation of
the Wilson-loop heavy quark-antiquark potential in the planar
limit of $\cN=4$ SYM to the classical minimal-surface problem of
calculating the action of a string connecting the quark and
antiquark on the boundary of ${\rm AdS}_5$ and extending into the
radial direction \cite{maldaloop}. However, when this method is
applied to more general backgrounds
\cite{wilsonloopTemp,bs,sonnenschein,sonnenschein2}, one often
encounters behaviors that are in sharp contrast with expectations
based on the gauge-theory side \cite{bs}. A possible resolution
could be that the parametric regions giving rise to these
behaviors are perturbatively unstable and hence unphysical.

\no Let us summarize the basic relevant facts for the case of
$\cN=4$ SYM. In the standard conformal case, dual to a stack of
D3-branes, this computation yields the expected Coulomb potential
and all fluctuations about the classical string configuration are
found to be stable \cite{cg,kmt}. In extensions to the theory at
finite temperature and at the Coulomb branch, dual to non-extremal
and multicenter \cite{trivedi,sfet1,warn,Basfe2} D3-branes
respectively, one expects to find a screened Coulomb potential
typical of thermal and Higgsed theories. However, one actually
encounters situations where (i) the potential has a second branch
of higher energy than the first, (ii) the potential exhibits a
confining behavior at large distances, and (iii) the screening
length is heavily dependent on the location of the probe string in
the internal space. Although these types of behavior are quite
counterintuitive, in a recent work \cite{stability1} we
established the reassuring result that the corresponding
parametric regions represent string configurations that are
perturbatively unstable, with the physical regions giving indeed
the expected screened Coulomb potential. Our analysis, partly
motivated by a mechanical analog, was based on the zero-mode
behavior of the differential equations governing the small
fluctuations about equilibrium and led to a formalism by means of
which instabilities may be found by exact or approximate analytic
methods. An important fact emerging from this analysis was the
existence of instabilities due to fluctuations of non-cyclic
angular coordinates, often appearing in cases where they are not
{\em a priori} expected. In particular, these fluctuations cast
the parametric region where confinement appears as unstable.

\no The analysis of \cite{stability1} was restricted to diagonal
metrics, where all fluctuations obey decoupled differential
equations. However, there are physically interesting situations
where the gravity duals where Wilson loops are evaluated contain
non-diagonal metrics. A first class of such backgrounds are
boosted (non-extremal) D3-brane backgrounds, used for evaluating
the potential for a quark-antiquark pair moving with respect to
the thermal medium \cite{psz,lrw,cgg,aev,cno,asz1,naok}. For this
case, the potential has a similar form to that in the zero-boost
case, namely it is a double-branched function of the length with
the lower branch corresponding to a screened Coulomb potential,
but the screening length falls off as
the velocity is increased. For this background, a numerical
stability analysis has been presented in \cite{michalogiorgakis},
indicating that the upper branch is perturbatively unstable. A
second class of such backgrounds are spinning non-extremal
D3-branes \cite{trivedi,cy,rs,spinningbranesthermo}, dual to
$\cN=4$ SYM at finite temperature and R-charge chemical
potentials. The behavior is similar to that in the absence of
chemical potentials, but the nontrivial dependence of the metrics
on certain angles again calls for a stability analysis. Finally, a
third class of such backgrounds are the Lunin--Maldacena
deformations \cite{LM} (see also
\cite{frolov,marginal-generalizations}) of D3-brane backgrounds,
dual to the Leigh--Strassler \cite{LS} deformations of $\cN=4$ SYM
which break supersymmetry down to $\cN=1$. These backgrounds are
characterized by the two real parameters $\g$ and $\s$. In the
usual situation where the quark and antiquark are not taken to be
separated in the internal space, only the $\s$--part of the
deformation affects the potential (see \cite{hsz} for
investigations on the effect of $\g$--deformations), resulting in
various behaviors ranging from the standard Coulomb behavior (for
deformed ${\rm AdS}_5 \times {\rm S}^5$) to complete screening and
linear confinement \cite{ahn-poritz,asz2} (for the deformed
multicenter D3-branes). To investigate the significance of these
results, a stability analysis is again in order. We note that now
the appearance of a confining potential is not in conflict with
expectations from the gauge-theory side.

\no This article is organized as follows: In section 2, we give a
brief review of the evaluation of Wilson loops in the backgrounds
under consideration. In section 3, we analyze small fluctuations
about the classical string configurations and we establish general
results which allow us to determine the regions of instability
using exact and approximate methods, generalizing the results of
\cite{stability1} to the non-diagonal case. In section 4, we apply
these results to the gravity duals under consideration and we
identify all unstable regions. In section 5, we summarize and
conclude. In appendix A, we present the complete list of angular
Schr\"odinger potentials for the backgrounds under consideration.
In appendix B, we present the detailed solution of the equation
for the angular fluctuations for a special case, in order to
provide a consistency check of the numerical calculation employed
in subsection 4.3.2. In appendix C, we apply our results to the
problem of open strings in Rindler space, which turns out to have
an exact classical-mechanical analog, namely the problem of the
shape of a soap film stretched between two circular rings.

\section{Wilson loops in AdS/CFT}
\label{sec-2}

In the framework of the AdS/CFT correspondence, the calculation of
the Wilson-loop potential of a heavy quark-antiquark pair proceeds
by considering a fundamental string whose endpoints lie on the two
temporal sides of the Wilson loop, and which extends in the radial
direction of the dual supergravity background so as to extremize
its worldsheet area. This type of calculation was first considered
in \cite{maldaloop} for the conformal case and was extended to
more general cases in
\cite{wilsonloopTemp,bs,sonnenschein,lrw,hsz}. Below, we give a
brief review of this procedure, adapted to the case where the
metric has off-diagonal elements in the directions transverse to
the quark-antiquark axis and the $B$--field has nonzero
components.

\no We consider a general background specified by a metric of the
form
\be
\label{2-1}
ds^2 = G_{tt} dt^2 + 2 G_{ti} dt dx_i + G_{ij} dx_i
dx_j + G_{yy} dy^2 + G_{uu} du^2 + G_{ab} d\th_a d\th_b + \ldots\
,
\ee
where, $y$ denotes the (cyclic) coordinate along the spatial side
of the Wilson loop, $u$ denotes the radial direction extending
from the UV at $u \to \infty$ down to the IR at some minimum value
$u_{\rm min}$ determined by the geometry, $x_i$ stands for a
generic cyclic coordinate, and $\th_a$ stands for a generic
non-cyclic coordinate. We also consider a $B$--field of the form
\be
\label{2-1a} B_2 = B_{ai} d \th_a \wedge d x_i \ ,
\ee
as is the case with many interesting gravity duals of gauge
theories \cite{LM,asz2,kt,ks,mn}. For future convenience, we introduce
the functions
\ba
\label{2-2} &g(u,\th_a) = - G_{tt} G_{uu}\ ,\qq f_y(u,\th_a) = -
G_{tt} G_{yy}\
, \nonumber\\
&f_{ij}(u,\th_a) = G_{ti} G_{tj} - G_{tt} G_{ij}\ ,\quad
f_{ab}(u,\th_a) = - G_{tt} G_{ab}\ , \quad h(u,\th_a) = G_{yy}
G_{uu}\ .
\ea

\no
According to AdS/CFT, the potential energy of the
quark-antiquark pair is given by
\be
\label{2-3} e^{-{\rm i} E T} = \langle W(C) \rangle = e^{{\rm i}
S[C]}\ ,
\ee
where $S[C]$ is the action for a string propagating in the
supergravity background whose endpoints trace the contour $C$. The
latter is given by the sum of Nambu--Goto and Wess--Zumino terms,
\be
\label{2-4} S[C] = - {1 \ov 2 \pi} \int d \tau d \sigma \left(
\sqrt{- \det g_{\a \b} } - \ha \epsilon^{\a\b} b_{\a\b} \right)\ ,
\ee
where $g_{\a\b}$ and $b_{\a\b}$ stand for the pullbacks
\be
\label{2-4a} g_{\a\b} = G_{\mu\nu}
\partial_\alpha x^\mu \partial_\b x^\nu \ , \qq b_{\a\b} = B_{\mu\nu}
\partial_\alpha x^\mu \partial_\b x^\nu\ .
\ee
To proceed, we employ the gauge fixing
\be
\label{2-5} t=\tau \ ,\qq u=\s \ ,
\ee
we assume translational invariance along $t$, and we consider the
radial embedding
\be
\label{2-6} y = y(u)\ ,\qq x_i =\const\ ,\qq \th_a = \th_{a 0} =
\const\ ,
\ee
supplemented by the boundary condition
\be
\label{2-7} u \left(\pm {L \ov 2} \right) = \infty\ ,
\ee
appropriate for a quark placed at $y=-L/2$ and an antiquark placed
at $y=L/2$. In the ansatz \eqn{2-7}, the constant values $\th_{a
0}$ of the non-cyclic coordinates $\th_a$ must be consistent with
the corresponding equation of motion. As we shall see later on,
this requires that
\be
\label{2-8}
\partial_a g(u,\th_a) |_{\th_a=\th_{a 0}} = \partial_a f_y(u,\th_a) |_{\th_a=\th_{a 0}} =
0\ ; \qq \partial_a \equiv {\partial \ov \partial \th_a}\ .
\ee
For the above ansatz, only the Nambu--Goto part of the action is
nonzero,\footnote{Nonzero contributions from the Wess--Zumino part
can arise when $B_{tu}$ and/or $B_{ty}$ are nonzero (see
\cite{sonnenschein2}). However, this occurs in a very restricted
class of backgrounds.} leading to
\be
\label{2-9} S = - {T \ov 2 \pi} \int d u \sqrt{ g(u) + f_y(u)
y^{\prime 2}}\ ,
\ee
where $T$ denotes the temporal extent of the Wilson loop, the
prime denotes a derivative with respect to $u$ while $g(u) \equiv
g(u,\th_{a 0})$ and $f_y(u) \equiv f_y(u,\th_{a 0})$. Conservation
of the momentum conjugate to $y$ implies that the classical
solution $y_{\rm cl}(u)$ satisfies
\be
\label{2-10}
y_{\rm cl}^\prime = \pm {\sqrt{f_{y0} F}\ov f_y}\ ,
\ee
where $u_0$ is the value of $u$ at the turning point, $f_{y0}
\equiv f_y(u_0)$, the two signs correspond to the two branches
around the turning point, and $F$ is defined as
\be
\label{2-11}
F = {g f_y \ov f_y - f_{y0}}\ .
\ee
Integrating \eqn{2-11}, we express the separation length as
\be
\label{2-12}
L = 2 f_{y0}^{1/2} \int_{u_0}^{\infty} d u {\sqrt{F} \ov f_y}\ ,
\ee
while inserting \eqn{2-10} into \eqn{2-9} and using \eqn{2-3}, we
obtain the potential energy
\be
\label{2-13}
E = {1 \ov \pi} \int_{u_0}^\infty d u \sqrt{F} - {1 \ov \pi}
\int_{u_{\rm min}}^\infty d u \sqrt{g}\ .
\ee
When the integrals \eqn{2-12} and \eqn{2-13} can be evaluated
exactly and the first one can be inverted for $u_0$, these
equations lead to an explicit expression for $E=E(L)$. However, in
practice this cannot be done, except for a few simple cases, and
Eqs. \eqn{2-12} and \eqn{2-13} are rather regarded as parametric
equations for $L$ and $E$ with parameter $u_0$.

\section{Stability analysis}
\label{sec-3}

We now turn to a stability analysis of these configurations just
discussed, our goal being to identify all parametric regions which
are unstable and for which information obtained using the
gauge/gravity correspondence cannot be trusted. In this section we
generalize the results of \cite{stability1} to non-diagonal
metrics of the form \eqn{2-1} and we establish a series of results
which will ultimately allow us to identify the stable and unstable
regions by a combination of exact and approximate analytic
methods.

\subsection{Small fluctuations}

To investigate the stability of the string configurations of
interest, we consider small fluctuations about the classical
solutions discussed above. In particular, we will be interested in
three types of fluctuations, namely (i) ``transverse''
fluctuations, referring to the cyclic coordinates $x_i$ transverse
to the quark-antiquark axis, (ii) ``longitudinal'' fluctuations,
referring to the cyclic coordinate $y$ along the quark-antiquark
axis, and (iii) ``angular'' fluctuations, referring to the
non-cyclic coordinates $\th_a$. The above fluctuations may be
parametrized by keeping the gauge choice \eqn{2-5} unperturbed and
perturbing the embedding as
\be
\label{3-1} x_i = \d x_i (t,u)\ ,\qq y = y_{\rm cl}(u) + \d y
(t,u)\ ,\qq \th_a = \th_{a 0} + \d \th_a(t,u)\ .
\ee
Inserting this ansatz in the action \eqn{2-4} and expanding in
powers of the fluctuations, we obtain the series
\be
\label{3-2} S = S_0 + S_1 + S_2 + \ldots\ ,
\ee
with the subscripts corresponding to the respective powers of the
fluctuations. The zeroth-order term gives just the classical
action. The first-order contribution reads
\ba
\label{3-3} \!\!\!\!S_1 = - {1 \ov 2\pi} \int dt du \left[
\sqrt{f_{y0}}\ \d y^\prime + \sqrt{{h F \ov g f_y}} G_{ti} \d
\dot{x}_i + \left( {1 \ov 2 F^{1/2}}
\partial_a g + {f_{y0} F^{1/2}
 \ov 2 f_y^2} \partial_a f_y \right) \d \th_a \right]\ ,
\ea
where $\partial_a g \equiv \partial_a g(u,\th_a) |_{\th_a=\th_{a
0}}$ and $\partial_a f_y \equiv \partial_a f_y(u,\th_a)
|_{\th_a=\th_{a 0}}$. The first term is a surface contribution
which is exactly cancelled by a similar term with the opposite
sign corresponding to the contribution of the linear order action
coming from the fluctuations of the lower string (cf. \eqn{2-10}),
provided we keep the variations of both strings at $u=u_0$ equal.
The second term is a total time derivative and hence is completely
irrelevant. Finally, in the third term, the coefficient of $\d
\th_a$ is just the equation of motion of $\th_a$ and the
requirement that it vanish leads indeed to the conditions
\eqn{2-8}. The second-order contribution is written as
\ba
\label{3-4} \!\!\!\!\!\!\!\!\!\!\!\! S_2 &=& - {1 \ov 2\pi} \int
dt du \biggl[ f_{ij} \left({1 \ov 2 F^{1/2}} \d x_i^{\prime} \d x_j^{\prime}
- {h F^{1/2} \ov 2 g f_y} \d \dot{x}_i \d \dot{x}_j \right) + {g f_y \ov 2 F^{3/2}}
\d y^{\prime 2} - {h \ov 2 F^{1/2}} \d \dot{y}^2 \nonumber\\
&& \qq\qq\quad\; - \sqrt{{h f_{y0} \ov g f_y}} G_{ti} ( \d y^{\prime} \d \dot{x}_i
- \d \dot{y} \d x^{\prime}_i ) + B_{ai} ( \d \th^{\prime}_a \d \dot{x}_i
- \d \dot{\th}_a \d x^{\prime}_i ) \\
&& \qq\qq\quad\; + f_{ab} \left( {1 \ov 2 F^{1/2}} \d
\th_a^{\prime} \d \th_b^{\prime} - {h F^{1/2} \ov 2 g f_y} \d
\dot{\th}_a \d \dot{\th}_b \right) \nonumber\\
&& \qq\qq\quad\; + \left( {1 \ov 4 F^{1/2}}
\partial_a \partial_b g
+ {f_{y0} F^{1/2} \ov 4 f_y^2} \partial_a \partial_b f_y \right)
\d \th_a \d \th_b \biggr]\ ,\nonumber
\ea
again with all functions and their $\th_a$--derivatives evaluated
at $\th_a=\th_{a 0}$. Writing down the equations of motion of the
fluctuations and introducing a harmonic time
dependence,
\be
\label{3-5} \d x^\m (t,u) = \d x^\m (u) e^{-{\rm i} \omega t}\ ,
\ee
we obtain the equations
\ba
\label{3-6} &&\!\!\!\!\!\!\!\!\!\!\!\! \left[{d \ov du}
\left({f_{ij} \ov F^{1/2}} {d \ov du} \right) + \omega^2 {h
F^{1/2} f_{ij} \ov g f_y} \right] \d x_j - {\rm i} \omega \left[
\partial_u \left( G_{ti} \sqrt{{h f_{y0} \ov g f_y}} \right) \d y
- \partial_u B_{ai} \d \th_a \right] = 0\ ,
\nonumber\\
&&\!\!\!\!\!\!\!\!\!\!\!\! \left[{d \ov du} \left( {g f_y \ov
F^{3/2}} {d \ov du} \right) + \omega^2 {h \ov F^{1/2}} \right] \d
y + {\rm i} \omega \partial_u \left( G_{ti} \sqrt{{h f_{y0} \ov g
f_y}} \right) \d x_i = 0\ ,
\\
&&\!\!\!\!\!\!\!\!\!\!\!\! \left[{d \ov du} \left( {f_{ab} \ov
F^{1/2}} {d \ov du} \right) + \omega^2 {h F^{1/2} f_{ab} \ov g
f_y}  - \left({1 \ov 2 F^{1/2}}
\partial_a \partial_b g + {f_{y0} F^{1/2} \ov 2 f_y^2} \partial_a
\partial_b f_y \right) \right] \d \th_b - {\rm i} \omega \partial_u B_{ai} \d x_i  =
0\ .\nonumber
\ea
for the transverse, longitudinal and angular fluctuations
respectively. We see that the fluctuations generically couple to
each other, satisfying a system of equations of the form
\be
\label{3-8}
\left\{ {d \ov du} \left[ \left( \begin{array}{ccc} \mathbf{p}_x & 0 & 0\\
0 & p_y & 0\\
0 & 0 & \mathbf{p}_\th\\
\end{array} \right) {d \ov du} \right] + \left( \begin{array}{ccc}
\omega^2 \mathbf{q}_x & -{\rm i} \omega \boldsymbol{\g} & {\rm i}
\omega \boldsymbol{\b}^T
\\ {\rm i} \omega \boldsymbol{\g}^T  & \omega^2 q_y & 0 \\ - {\rm i}
\omega \boldsymbol{\b} & 0 & \omega^2 \mathbf{q}_\th +
\mathbf{r}_\th\end{array}
\right) \right\} \left( \begin{array}{c} \d \mathbf{x} \\ \d y \\
\d\boldsymbol{\th}
\end{array} \right) = 0\ ,
\ee
where the matrices $\mathbf{p}_x = ( p_{ij} )$, $\mathbf{p}_\th =
( p_{ab} )$, $\mathbf{q}_x = ( q_{ij} )$, $\mathbf{q}_\th = (
q_{ab} )$, $\mathbf{r}_\th = ( r_{ab} )$, $\boldsymbol{\b} =
(\b_{ai})$ and the column vector $\boldsymbol{\g}=(\g_i)$ are read
off from Eqs. \eqn{3-6}. The problem is defined in the interval
\be
\label{3-10} u_0\leqslant u < \infty\ ,\qq u_0 > u_{\rm min}\ ,
\ee
and the boundary conditions for the fluctuations, determined by
requiring that the string endpoints be held fixed and that the two
parts of the string glue smoothly at $u=u_0$ read
\ba
\label{3-11} &&\lim_{u\to \infty} \d x_i(u)=0\ ,\qq \lim_{u\to
u_0^+} (u-u_0)^{1/2} \d x_i^\prime(u) = 0\ ,
\nonumber\\
&&\lim_{u\to \infty} \d y(u)=0\ ,\qq \lim_{u\to u_0^+}
\left[ \d y(u) + 2 (u-u_0) \d y^\prime(u) \right] = 0\ , \\
&&\lim_{u\to \infty} \d \th_a (u)=0\ ,\qq \lim_{u\to u_0^+}
(u-u_0)^{1/2} \d \th_a^\prime(u) = 0\ . \nonumber
\ea
These boundary conditions follow from a straightforward extension
of the proof given in \cite{stability1} and is essentially based
on the fact that $u=u_0$ and $u=\infty$ present regular singular
points of the system \eqn{3-6}. Therefore, our stability analysis
has reduced to a coupled system of Sturm--Liouville equations,
with our objective being to determine the range of values of $u_0$
for which the lowest eigenvalue becomes negative.

\no
Ideally, one would like to solve the above Sturm--Liouville
equations exactly, obtain the lowest eigenvalue $\omega_0^2$ in
terms of the parameter $u_0$ and determine the regions where
$\omega_0^2$ becomes negative. However, in most cases, this is
impossible due to the complexity of the equations. On the other
hand, it turns out that we can obtain useful information by
studying a simpler problem, namely the zero-mode problem of the
associated differential operators. Regarding the transverse and
longitudinal fluctuations, we will use our Sturm--Liouville
description to prove that transverse zero modes do not exist while
longitudinal zero modes are in one-to-one correspondence with the
critical points of the function $L(u_0)$. For the angular
fluctuations, we will employ an alternative Schr\"odinger
description which allows us to identify angular zero modes using
either exact or approximate methods.

\subsection{Zero modes}

To motivate the significance of zero modes for our stability
analysis, we consider a Sturm--Liouville system of the form
considered earlier on and we let $\omega_n^2(u_0)$ be the
corresponding eigenvalues. By standard results of Sturm--Liouville
theory, these eigenvalues are real and strictly-ordered in the
sense that $\omega_0^2(u_0) < \omega_1^2(u_0) < \ldots$, from
which it also follows that different eigenvalues do not cross as
we vary $u_0$. Furthermore, we know that, for sufficiently large
values of $u_0$, all eigenvalues are positive. The above
considerations imply that the first occurrence of instabilities
will arise at a value of $u_0$ at which the lowest eigenvalue
$\omega_0^2(u_0)$ becomes negative. Therefore, a necessary
condition for the appearance of instabilities is the existence of
values $u_{0 \rm c}^{(i)}$ of $u_0$ for which $\omega_0^2(u_0)=0$.
To verify that these zero modes really correspond to points where
$\omega_0^2$ changes sign, we may solve the full Sturm--Liouville
system near each critical point $u_{0 \rm c}^{(i)}$, a task that
may be accomplished using perturbative \cite{stability1} or
numerical methods. In fact, as we shall see later on, in all
examples considered in this paper, a zero mode always signifies a
change of sign of $\omega_0^2$ i.e. marks the boundary between a
stable and an unstable region.

\no Restricting to zero modes simplifies our problem tremendously,
as the three equations in \eqn{3-6} decouple, reducing to
\ba
\label{3-12}
&&{d \ov du} \left({f_{ij} \ov F^{1/2}} {d \ov du} \d
x_j \right) = 0\ ,\nonumber\\
&&{d \ov du} \left( {g f_y \ov F^{3/2}} {d \ov du} \d y \right) =
0\ ,\\
&&\left[{d \ov du} \left( {f_{ab} \ov F^{1/2}} {d \ov du} \right)-
\left({1 \ov 2 F^{1/2}} \partial_a \partial_b g + {f_{y0} F^{1/2}
\ov 2 f_y^2} \partial_a
\partial_b f_y \right) \right] \d \th_b =
0\ .\nonumber
\ea
Using these simplified expressions, we will next prove that transverse
zero modes do not exist and that longitudinal zero modes are in
one-to-one correspondence with the critical points of the function
$L(u_0)$, and we will devise analytic methods for seeking angular
zero modes. This is a fairly straightforward extension of the
results of \cite{stability1}.

\subsubsection{Transverse zero modes}

\no We consider first the case of the transverse fluctuations.
Using the definition of $F$ in \eqn{2-11}, performing an
integration by parts, and expanding about $u=u_0$, we write the
general solution of the first of \eqn{3-12} as
\ba
\label{3-13} \d x_i &=& \int^{\infty}_u du \sqrt{gf_y\ov
f_y-f_{y0}} f^{-1}_{ij} c_j
\nonumber\\
& = & -2 {\sqrt{g f_y}\ov f_y^\prime} \sqrt{f_y-f_{y0}}
f^{-1}_{ij} c_j - 2 \int_u^\infty du \sqrt{f_y-f_{y0}} \del_u
\left({\sqrt{g f_y}\ov f_y^\prime} f^{-1}_{ij}\right) c_j
\\
& = &  - 2 \int_{u_0}^\infty du \sqrt{f_y-f_{y0}} \ \del_u
\left({\sqrt{g f_y}\ov f_y^\prime} f^{-1}_{ij} \right) c_j
  - 2 \sqrt{g_0 f_{y0}\ov f^\prime_{y0}} f^{-1}_{ij0} c_j (u-u_0)^{1/2} + {\cal O}(u-u_0)\ ,
\nonumber
\ea
where the $c_i$ are constants. In order for this zero mode to
exist, it must satisfy the first boundary condition in \eqn{3-11},
which requires the coefficient of $(u-u_0)^{1/2}$ to vanish. As
the matrix $f^{-1}_{ij0}$ turns out not to admit any nontrivial
null eigenvectors,\footnote{We have not a general proof of that
statement, but this is indeed the case in all examples of the
present paper as well as of \cite{stability1}.} this is only
achieved when all $c_i=0$ i.e. for the trivial solution $\d
x_i=0$. Therefore, the transverse fluctuations do not lead to zero
modes of the system \eqn{3-6}.

\subsubsection{Longitudinal zero modes}

\no Turning to the longitudinal fluctuations, the fact that they
decouple from the rest in the zero-mode analysis implies that the
results of \cite{stability1} for the diagonal case hold in our
case as well. In particular, in \cite{stability1} it was proven
that the longitudinal zero-mode solution can be expressed in terms
of the $u_0$--derivative of the length function as follows
\be
\label{3-14} \d y \sim 2 \sqrt{g_0 f_{y0}\ov f_{y0}^{\prime 3}}\
(u-u_0)^{-1/2} + {\sqrt{f_{y0}} \ov f_{y0}^\prime} L^\prime (u_0)
 + {\cal O}\left((u-u_0)^{1/2}\right)\ .
\ee
In order for this zero mode to exist, it must satisfy the second
boundary condition in \eqn{3-11}, which requires the constant term
to vanish. Therefore, the solution exists only if
\be
\label{3-15} L^\prime (u_0) = 0\ ,
\ee
or, equivalently, if \cite{stability1}
\be
\label{3-15a} \int_{u_0}^\infty { du\ov \sqrt{f_y-f_{y0}}} \
\partial_u \left(\sqrt{g f_y}\ov f_y^\prime \right) = 0  \ ,
\ee
\begin{figure}[!t]
\begin{center}
\begin{tabular}{cccc}
 \includegraphics[height=2.9cm]{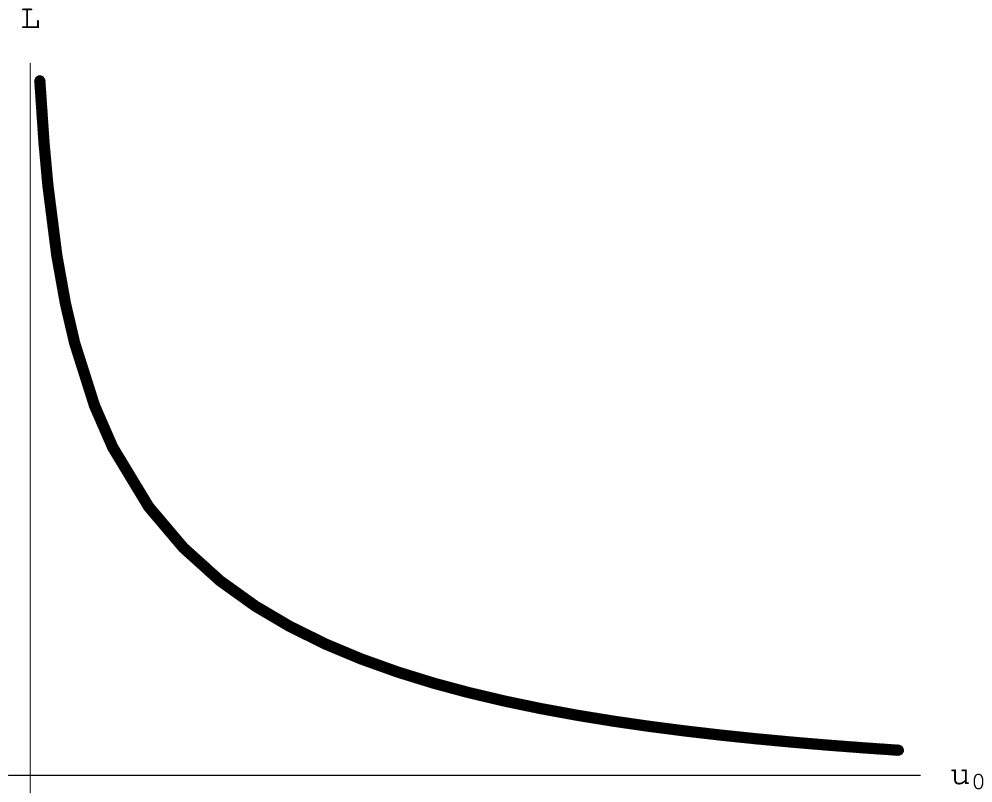}
&\includegraphics[height=2.9cm]{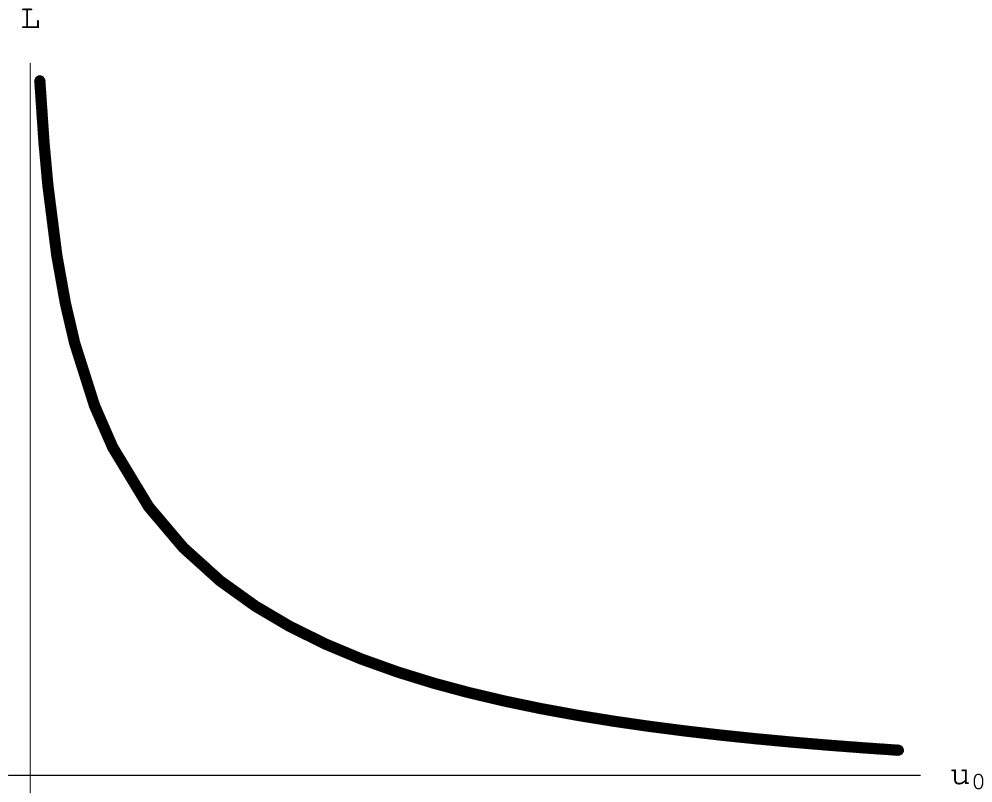}
&\includegraphics[height=2.9cm]{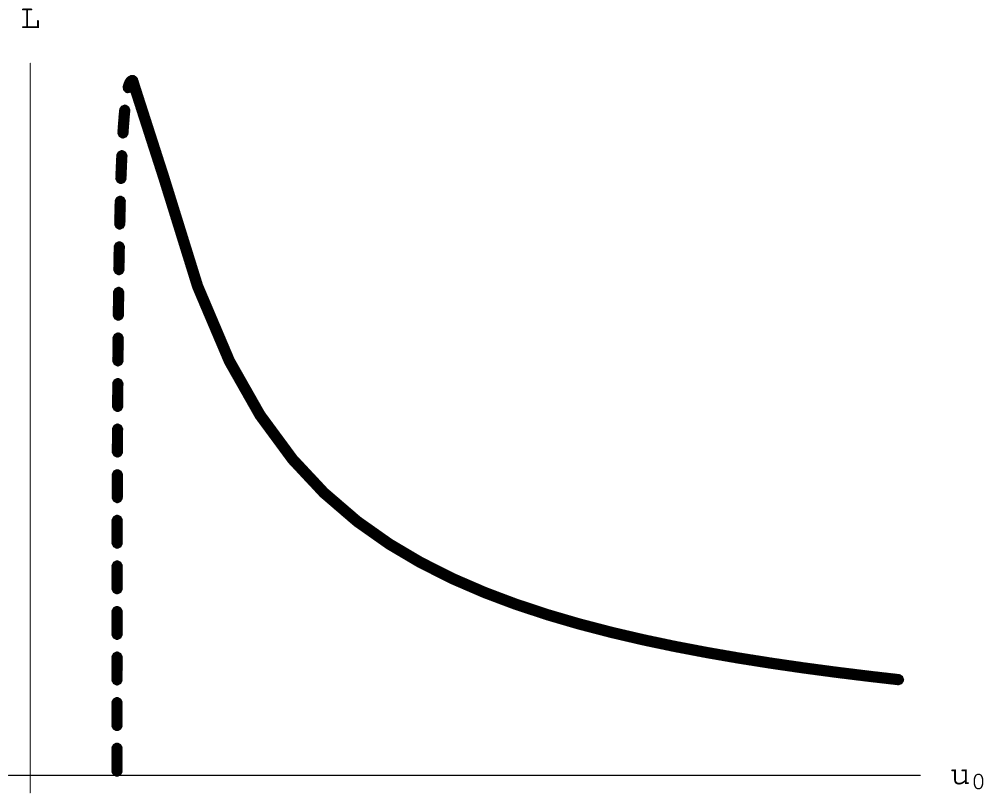}
&\includegraphics[height=2.9cm]{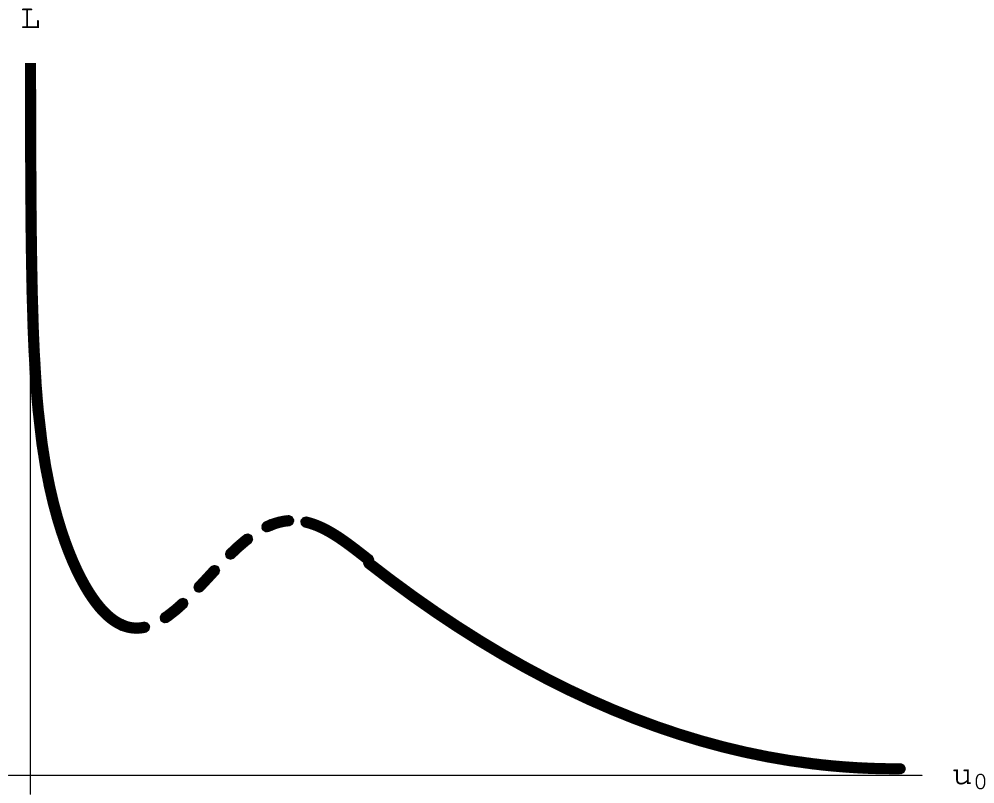} \\
 \includegraphics[height=2.9cm]{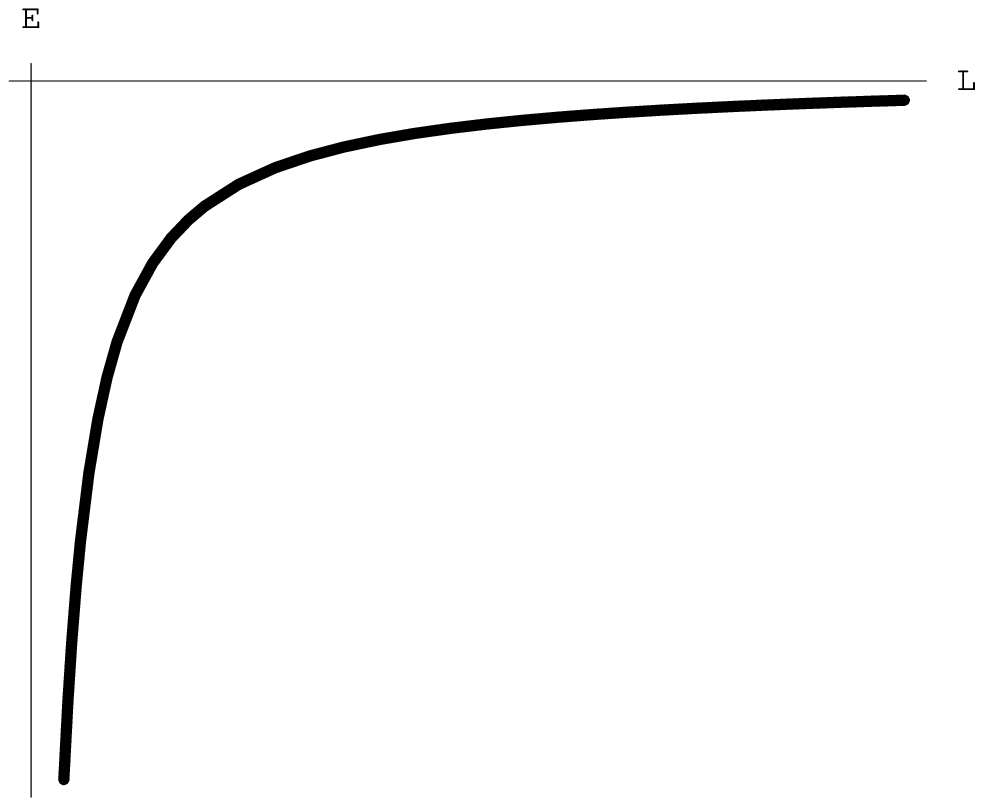}
&\includegraphics[height=2.9cm]{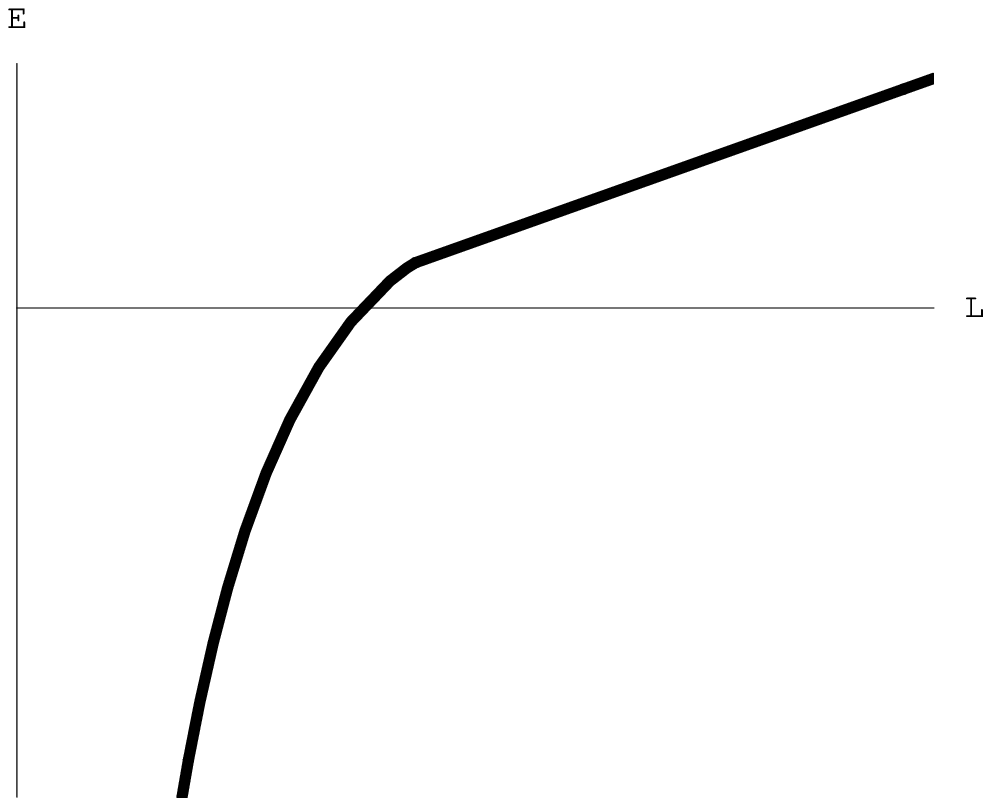}
&\includegraphics[height=2.9cm]{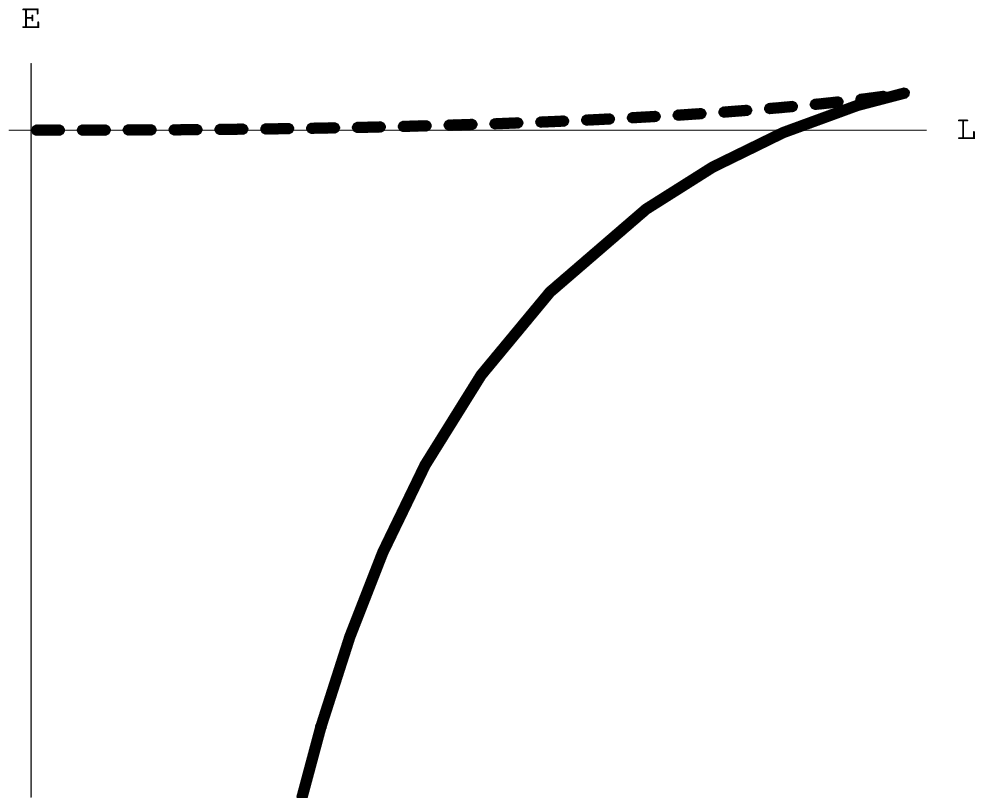}
&\includegraphics[height=2.9cm]{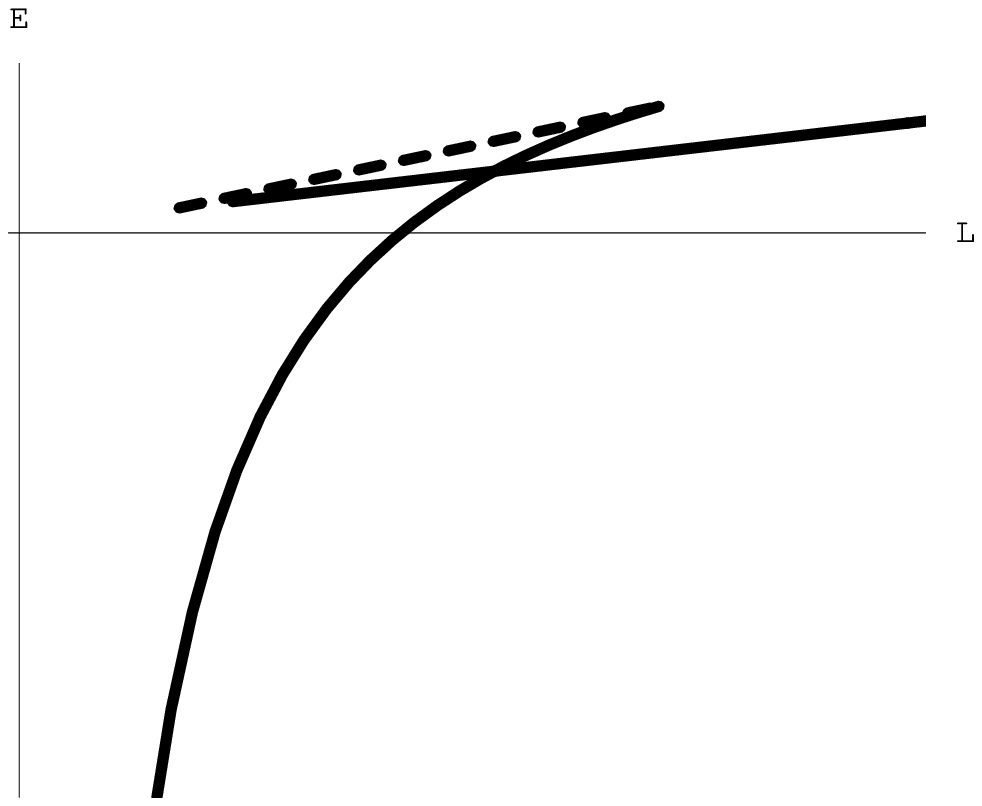} \\
(a) & (b) & (c) & (d)
\end{tabular}
\end{center}
\caption{The four generic types of behavior of $L(u_0)$ and $E(L)$
encountered in AdS/CFT calculations of Wilson loops, corresponding
to (a) no critical points, Coulomb or screened Coulomb potential,
(b) no critical points, Coulomb/confining potential, (c) one
critical point, double-branched potential and (d) two critical
points, multi-branched potential. The dashed parts of the curves
indicate regions that are unstable under longitudinal
perturbations.
} \label{fig1}
\end{figure}which requires that the derivative term should change sign at least
once as $u$ ranges in the interval $u\in [u_0,\infty)$. Therefore,
given the critical points of the length function, we may determine
all values $u_{0 \rm c}$ of $u_0$ where the system \eqn{3-6} has a
zero mode. In view of our earlier remark that zero modes always
signify a transition from a stable to an unstable region in our
examples, the longitudinal instabilities for the various
configurations of interest are as shown in Fig.~1. Additional
instabilities may arise from angular perturbations, so that
part(s) of the curves, stable under longitudinal perturbations
only, might be actually unstable.

\subsubsection{Angular zero modes}

We finally consider angular fluctuations, which must be treated
separately due to the fact that the non-cyclicity of the angular
coordinates induces a ``restoring force'' term in the
corresponding Sturm--Liouville equation, implying that we cannot
write $\d \th_a$ in an explicit integral form, even in the
zero-mode case. On the other hand, if we restrict to situations
where the angular fluctuations are decoupled from the rest (which
is indeed the case in most of our examples), the equation
satisfied by each of the fluctuations can be easily brought to a
Schr\"odinger form. This description allows us to determine the
zero modes to quite high accuracy by using approximate methods.
Following \cite{stability1}, we outline the procedure below.

\no Letting $\th$ be any of the decoupled non-cyclic angular
variables, the Sturm--Liouville equation for its fluctuations
reads
\be
\label{3-16} \left[ - {d \ov du} \left( p_\th {d \ov du} \right) -
r_\th \right] \d \th  = \omega^2 q_\th \d \th\ .
\ee
Employing the change of variables
\be
\label{3-17} x = \int_u^\infty du' \sqrt{q_\th\ov p_\th}\ \ ,\qq
\d\th = (p_\th q_\th)^{-1/4} \Psi\ ,
\ee
we may transform Eq. \eqn{3-16} to a standard Schr\"odinger
equation
\be
\label{3-18} \left[ -{d^2 \ov dx^2} + V_\th (x;u_0) \right]
\Psi(x) = \omega^2 \Psi(x)\ ,
\ee
with the potential
\be
\label{3-19} V_\th = -{r_\th \ov q_\th} + {p_\th^{1/4} \ov
q_\th^{3/4}} {d \ov du} \left[ \left( {p_\th \ov q_\th}
\right)^{1/2} {d \ov du}  (p_\th q_\th)^{1/4} \right] =  -{r_\th
\ov q_\th} + (p_\th q_\th)^{-1/4} {d^2 \ov dx^2} (p_\th
q_\th)^{1/4}\ .
\ee
The problem is defined in the interval
\be
\label{3-20} 0 \leqslant x \leqslant x_0\ ,\qq x_0 =
\int_{u_0}^\infty du \sqrt{q_\th \ov p_\th}\ ,
\ee
which is finite, making the fact that the fluctuation spectrum is
discrete manifest. The boundary conditions are simply
\be
\label{3-21} \Psi(0) = 0 \ ,\qq \Psi^\prime(x_0)= 0 \ .
\ee
Note that, in general, Eq. \eqn{3-17} does not lead to a closed
expression for $u$ in terms of $x$ and therefore it is not always
possible to write down the potential as an explicit function of
$x$. In what follows, we will encounter cases where $V_\th(x)$ can
be determined exactly as well as cases where it may be adequately
approximated by solvable potentials.

\no To develop our approximate methods, we first note that the
behavior of the angular Schr\"odinger potential $V_\th(u;u_0)$ in
the limits $u \to \infty$ and $u=u_0$ is given by
\be
\label{3-22} V_\th(\infty,u_0)=a u^2+V_\infty  ,
\ee
and
\be
\label{3-23} V_\th(u_0;u_0)  = {1 \ov 8} {g_0 f_{y0} f_{y0}^\prime
\ov h_0^2 f_{\th 0}^2}\ \del_{u_0} \left(h_0 f_{\th 0}^2\ov g_0
f_{y0}\right) + \ha {g_0\ov h_0 f_{\th 0}} \del^2_\th f_{y0}
\equiv V_0\ ,
\ee
where $a$, $V_\infty$ and $V_0$ are finite quantities that depend
on the parameter $u_0$. When $a \geqslant 0$, the potential
expressed in terms of the variable $x$ rises from a minimum value
to infinity in the finite interval $x\in [0,x_0]$,
and hence it is reasonable to approximate it
by an infinite well given by
\be
\label{3-24}
V_{\rm approx} = \left\{ \begin{array}{cl} \overline{V} , & \quad 0\leqslant x \leqslant x_0\\
\infty\ , & \quad {\rm otherwise} \end{array}\right\}\ ;\qq
\overline{V} \equiv \left\{ \begin{array}{cl} \ha( V_0+V_\infty), & \quad a = 0\\
V_0 , & \quad a \ne 0 \end{array}\right\}\ .
\ee
With the boundary conditions \eqn{3-19} the energy levels read
\be
\label{3-25} \omega_n^2\  = {(2n+1)^2 \pi^2 \ov 4[x_0(u_0)]^2} +
\overline{V}\ .
\ee
When $a=0$, this approximation is valid for both the ground state
and the excited states. When $a > 0$ however, the approximation is
valid only for the ground state as excited states are affected by
the details of the exact potential. In any case, we may determine
the critical value $u_{0 \rm c}$ by solving the equation
$\omega^2_0(u_0)=0$; clearly, a solution to this equation exists
only if the average $\overline{V}$ is negative at least in a
finite range of values of $u_0$. The infinite-well approximation
just described is to be used for obtaining an indication for the
existence of zero modes and for estimating $u_{0 \rm c}$. In most
cases, these estimates are quite close to the exact values, determined
through numerical analysis.

\section{Applications}
\label{sec-4}

Having developed the necessary formalism, we now turn to analyzing
the stability properties of the string configurations of interest
in backgrounds of boosted non-extremal D3-branes, spinning
D3-branes and marginally-deformed D3-branes, all of which fall in
the general category described by the metrics \eqn{2-1}. For each
case under consideration, we give the formulas for the length and
the potential energy and we perform a stability analysis according
to the guidelines of the previous section. Depending on the
problem at hand, the regions of stability are determined by exact
or approximate analytic methods, by the numerical solution of
certain algebraic or transcendental equations, or by the numerical
evaluation of certain integrals. In all cases, this represents a
considerable improvement, both conceptual as well as practical,
over the direct numerical solution of the
differential equations governing the fluctuations.

\subsection{Boosted non-extremal D3-branes}

The first type of metrics we will consider are obtained by a boost
of the metric for non-extremal D3-branes along one brane direction
transverse to the quark-antiquark axis, say $x$. They are given by
\ba
\label{4-1} ds^2 &=& {u^2\ov R^2} \left[-\left(1-{\g^2 \m^4 \ov
u^4}\right) dt^2 + 2{\g^2 v \m^4 \ov u^4} dt dx + \left(1+{\g^2
v^2 \m^4 \ov u^4}\right) dx^2 + dy^2 +
 dz^2\right]\nonumber\\
&+& R^2 \left( {u^2 \ov u^4-\m^4}\ du^2 + d\Om_5^2 \right)\ ,
\ea
where $v$ is the boost velocity and $\g=1/\sqrt{1-v^2}$.  The
metric has a horizon at
\be
\label{4-2} u_H = \m\ ,
\ee
while there also exists a velocity-dependent radius $u_\g =
\sqrt{\g} \m \geqslant u_H$ beyond which the string cannot
penetrate (see \cite{psz,lrw,cgg,aev,asz1} for discussions). The
Hawking temperature of the solution is given by $T=\m /\pi R^2$.
For the stability analysis for this metric, the ``transverse''
coordinates are $(x,z,\Omega_5)$, with $x$ coupled to the
longitudinal coordinate $y$, while no ``angular'' coordinates
exist. This latter fact implies that the position of the string in
the internal space can be arbitrary.

\no The calculation of the Wilson loops of section 2 for the above
metric yields the potential for a quark and antiquark moving with
velocity $v$ with respect to a thermal plasma at a temperature $T$
and may be used as a crude model for the study of meson
dissociation in plasmas. Switching to dimensionless units by the
change of variables
\be
\label{4-3} u \to \m u\ ,\qq u_0 \to \m u_0\ ,\qq L \to {R^2 \ov
\m} L \ ,\qq E \to {\m \ov \pi} E\ ,
\ee
we find that the length and the potential energy read
\cite{lrw,asz1}
\be
\label{4-4} L(u_0,\g) = {2 \sqrt{2} \pi^{3/2} \ov \Gamma(1/4)^2}\
{\sqrt{u_0^4-\g^2} \ov u_0^3}\ {}_2F\!_1 \left({1 \ov 2},{3 \ov
4},{5 \ov 4};{1 \ov u_0^4} \right)\ ,
\ee
and
\be
\label{4-5} E(u_0,\g) =  - {\sqrt{2} \pi^{3/2} \ov \Gamma(1/4)^2}
\left[ u_0 \, {}_2F\!_1\left(-{1 \ov 4},{1 \ov 2},{1 \ov 4};{1 \ov
u_0^4} \right) + {\g^2 \ov u_0^3}\ {}_2F\!_1\left({1 \ov 2},{3 \ov
4},{5 \ov 4};{1 \ov u_0^4} \right) \right] + 1\ ,
\ee
where ${}_2F\!_1(a,b,c;x)$ is the standard hypergeometric
function. The behavior of the length and the energy is as in
Fig.~1c, with the maximal value of the length, $L_{\rm
c}(u_0,\g)$, being a decreasing function of the velocity and
satisfying the approximate law $L_{\rm c}(u_0,\g) \simeq \g^{-1/2}
L_{\rm c}(u_0,1)$ \cite{lrw} indicating an enhancement of the
dissociation rate of the quark-antiquark bound state with
increasing velocity.

\begin{figure}[!t]
\begin{center}
\begin{tabular}{c}
\includegraphics[height=6cm]{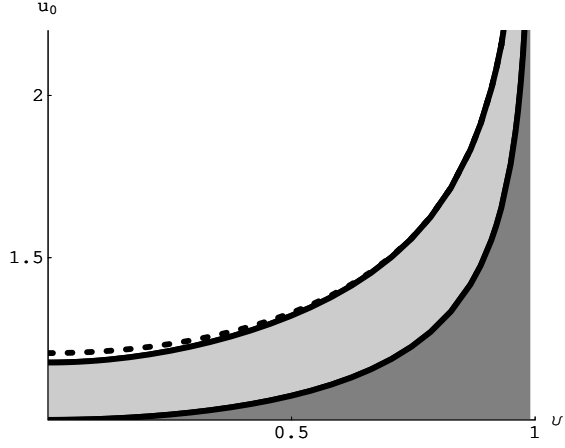}
\end{tabular}
\end{center}
\vskip -.5 cm \caption{Stability diagram in the $(v,u_0)$ plane
for the case of the boosted non-extremal D3-brane background. The
light-shaded region corresponds to instabilities under
longitudinal perturbations and the dotted curve indicates our
approximation \eqn{4-6a} for $u_{0 \rm c}(\g)$. The dark-shaded
region corresponds to the values $u_0 < \sqrt{\g}$ which are
inaccessible to the string. The region $u_0 < 1$ beyond the
horizon has been excluded from the diagram.} \label{fig2}
\end{figure}

\no Turning to the stability analysis, our general results imply
that the coupled system of the $(\d y,\d x)$ fluctuations has
instabilities for $u_0 < u_{0 \rm c}(\g)$ where $u_{0 \rm c}(\g)$
is the critical point of $L(u_0,\g)$. The latter is determined by
the solution of \eqn{3-15} which leads to the transcendental equation
\be
\label{4-6} 5 u_0^4(u_0^4-3 \g^2) \, {}_2F\!_1 \left({1 \ov 2},{3
\ov 4},{5 \ov 4},{1 \ov u_0^4}\right) + 6 (u_0^4- \g^2) \,
{}_2F\!_1 \left({3 \ov 2},{7 \ov 4},{9 \ov 4},{1 \ov u_0^4}\right)
= 0\ ,
\ee
which can be obtained numerically. For the physically interesting
case $\g \gg 1$ (in which we also have $u_0 \gg 1$) we can solve
this equation perturbatively in $1/\g$ with the result
\be
\label{4-6a} u_{0 \rm c}(\g) = 3^{1/4}\sqrt{\g}\left[ 1-{1\ov
15\g^2}-{23\ov 1350\g^4} + \cO(\g^{-6}) \right] .
\ee
The results of the stability analysis just presented are
summarized in the diagram of Fig.~2. Note that the estimate
\eqn{4-6a} for $u_{0 \rm c}(\g)$ is remarkably close to the exact
numerical result even for low velocities. Our results are in
accordance with those of the numerical analysis presented in
\cite{michalogiorgakis}.

\no The above analysis may be extended to the case where the
velocity and the axis of the quark-antiquark pair are not
perpendicular but form an angle smaller than $\pi/2$. For such a
case, we may modify our ansatz \eqn{2-6} to include a variable $x$
that has non-trivial dependence on $u$, while in the special case
of motion parallel to the axis we may equivalently boost along the
$y$ instead of the $x$ axis which would mean that the metric
\eqn{2-1} would include an extra $G_{ty}$ term. Such string
solutions were considered in \cite{lrw,aev,asz1,naok} and the
behavior of the length and energy is again as in Fig.~1c. Hence we
expect instabilities under longitudinal perturbations for $u_0 <
u_{0 \rm c}(\g)$, a fact that we have actually verified by a
small-fluctuation analysis which is however too lengthy to be
included here.

\subsection{Spinning D3-branes}

We next consider the case of spinning (non-extremal) D3-branes,
dual to $\cN=4$ SYM theory at finite temperature and R-charge
chemical potentials. These metrics were found in full generality
in \cite{trivedi}, based on previous results from \cite{cy}, and
their thermodynamical properties were examined in
\cite{spinningbranesthermo}. In the conventions of \cite{rs} which
we here follow, the field-theory limit of these solutions is
characterized by the non-extremality parameter $\m$ and the
angular momentum parameters $a_i$, $i=1,2,3$. Here, we will
restrict to two special cases, corresponding to two equal nonzero
angular momenta, $a_2=a_3=r_0$, and one nonzero angular momentum,
$a_1=r_0$, to which we will apply our stability analysis.

\subsubsection{Two equal nonzero angular momenta}

For the case of two equal angular momenta, the metric reads
\ba
\label{4-7} ds^2 &=& H^{-1/2} \left[-\left(1-{\m^4 H \ov
R^4}\right) dt^2 + dx^2 + dy^2 + dz^2\right] + H^{1/2}
{u^4(u^2-r_0^2 \cos^2\th)\ov (u^4-\m^4)(u^2-r_0^2)}\ du^2
\nonumber\\
&+& H^{1/2}\Big[(u^2-r_0^2\cos^2\th )d\th^2 + u^2 \cos^2\th
d\Om_3^2 + (u^2-r_0^2)\sin^2\th d \phi_1^2
\\
&& \qq\quad - \:\:  2 {\m^2 r_0\ov R^2} \ dt \cos^2\th (\sin^2\psi
d \phi_2 + \cos^2\psi d \phi_3)\Big]\ , \nonumber
\ea
where
\be
\label{4-8} H={R^4\ov u^2 (u^2-r_0^2\cos^2 \th)} \
\ee
and $d\Om^2_3 = d\psi^2 + \sin^2 \psi d \phi_2^2 + \cos^2 \psi d
\phi_3^2$. The metric has a horizon at
\be
\label{4-9} u_H = \m\ ,
\ee
and considerations of thermodynamic stability restrict the ratio
$\l \equiv \m/r_0$ according to
\be
\label{4-10} \l \geqslant 1 \hbox{ (CE)}\ ,\qq \l \geqslant
\sqrt{2} \hbox{ (GCE)}\ ,
\ee
where the two values refer to the canonical and grand-canonical
ensembles respectively, with the former giving the lowest bounds
in our examples.\footnote{For details on the thermodynamic
stability of spinning branes see \cite{spinningbranesthermo}. For
a brief summary for the type of the spinning D3-branes used in
this paper see \cite{AS1}.} For the stability analysis for this
metric, the ``transverse'' coordinates are
$(x,z,\psi,\phi_{1,2,3})$, with $\phi_{2,3}$ coupled to the
longitudinal coordinate $y$, and the only ``angular'' coordinate
is $\th$. The restriction \eqn{2-8} on the angular location
$\th_0$ of the string allows only trajectories with $\th_0=0$ and
$\th_0=\pi/2$. To keep the discussion to a reasonable length, we
will here restrict to the $\th_0=\pi/2$ trajectories. For the
trajectories with $\th_0=0$, the longitudinal and angular
fluctuations may be analyzed in the same way, using the
Schr\"odinger potentials given in appendix A for the latter case.

\no To examine the Wilson-loop computation, we switch to
dimensionless units by setting
\be
\label{4-11} u \to r_0 u\ ,\qq u_0 \to r_0 u_0\ ,\qq L \to {R^2
\ov r_0} L \ ,\qq E \to {r_0 \ov \pi} E\ .
\ee
Then, we find that the length and the potential energy are given
by the expressions
\be
\label{4-12} L(u_0,\l) = 2 \sqrt{u_0^4-\l^4} \int_{u_0}^\infty {du
\, u \ov \sqrt{(u^2-1)(u^4-u_0^4)(u^4-\l^4)}}\ ,
\ee
and
\be
\label{4-13} E(u_0,\l) = \int_{u_0}^\infty du \, u \sqrt{u^4-\l^4
\ov (u^2-1)(u^4-u_0^4)} - \int_{u_H}^\infty du {u \ov
\sqrt{u^2-1}}\ ,
\ee
which unfortunately cannot be evaluated in closed
form.\footnote{Such integrals can be thought of as periods of
Riemann surfaces corresponding to algebraic curves. In this case
the genus of the Riemann surfaces is at least two.} Their behavior
is as in Fig.~1c.

\begin{figure}[!t]
\begin{center}
\begin{tabular}{c}
\includegraphics[height=6cm]{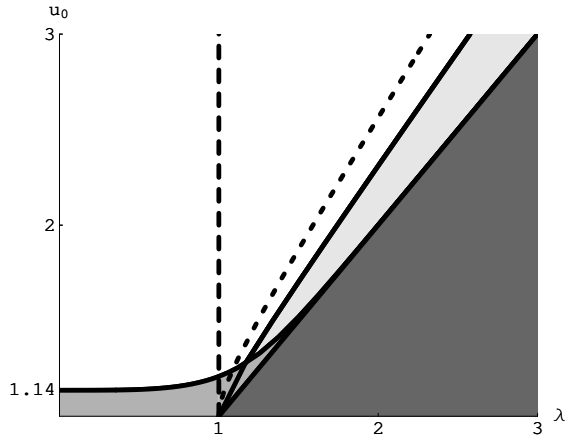}
\end{tabular}
\end{center}
\vskip -.5 cm \caption{Stability diagram in the $(\l,u_0)$ plane
for the spinning D3-brane with two angular momenta and for the
trajectory with $\th_0=\pi/2$. The two light-shaded regions on the
upper right-hand and lower left-hand sides correspond to
instabilities under longitudinal and angular perturbations,
respectively. The dark-shaded region corresponds to values of
$u_0$ beyond the horizon. The region to the left of the vertical
dashed line corresponds to the values $\l<1$ for which the
background is thermodynamically unstable. The dotted line
corresponds to the curve drawn using the approximation described
below \eqn{4-14}.} \label{fig3}
\end{figure}

\no Turning to the stability analysis, our general results imply
that the coupled system of the $(\d y,\d \phi_{2,3})$ fluctuations
has instabilities for $u_0 < u_{0 \rm c}^{(y)}(\l)$ where $u_{0
\rm c}^{(y)}(\l)$ is the critical point of $L(u_0,\l)$. This value
can be obtained by numerically solving the equation \eqn{3-15a}
which after some algebra takes the simple form
\be
\label{4-14} \int_{u_0}^\infty du { u^6-3\l^4 u^2 + 2 \l^4 \ov u^3
\sqrt{(u^2-1)^3 (u^4-u_0^4)(u^4-\l^4)}} = 0\ .
\ee
In order for a solution to exist, the integrand must change sign
at least once in the interval $u\in [u_0,\infty)$. This
requirement leads to the inequality
$\l<u_0<\l\sqrt{2\cos(\phi/3)}$, where $\cos\phi=-1/\l^2$, hence
$\phi\in[\pi/2,\pi]$. Taking the upper bound as an equality, i.e.
$u_{0 \rm c}^{(y)}(\l)= \l\sqrt{2\cos(\phi/3)}$, gives a quite
good approximation to the exact numerical result, as one can see
by inspection of Fig.~3.

\no Finally, for the angular fluctuations, the Schr\"odinger
potential reads
\be
\label{4-15} V_\th = -1\ ,
\ee
implying that the infinite-well approximation is exact in this
case. Therefore, the critical value of $u_0$ beyond which
instabilities occur are obtained by using Eq. \ref{3-25}, which is
in fact exact in this case, with $\overline{V}=-1$ and solving the
equation $\omega_0^2=0$. The resulting equation has the form
\be
\label{4-16} \int_{u_0}^\infty {du \, u^3 \ov
\sqrt{(u^2-1)(u^4-u_0^4)(u^4-\l^4)}} = {\pi \ov 2}\ ,
\ee
and again can be solved numerically to give $u_{0 \rm
c}^{(\th)}(\l)$. The results of this stability analysis are
summarized in the diagram of Fig.~3.

\subsubsection{One nonzero angular momentum}

For the case of one angular momentum, the metric reads
\ba
\label{4-17} ds^2 &=& H^{-1/2} \left[-\left(1-{\m^4 H \ov
R^4}\right) dt^2 + dx^2 + dy^2 + dz^2\right] + H^{1/2}
{u^2(u^2+r_0^2 \cos^2\th)\ov u^4+r_0^2 u^2 -\m^4}\ du^2
\nonumber\\
&+& H^{1/2}\Big[(u^2+r_0^2\cos^2\th )d\th^2 + u^2 \cos^2\th
d\Om_3^2 + (u^2+r_0^2)\sin^2\th d \phi_1^2
\\
&& \qq\quad - \:\: 2 {\m^2 r_0\ov R^2} \sin^2\th dt d\phi_1\Big]\
, \nonumber
\ea
where
\be
\label{4-18} H={R^4\ov u^2 (u^2+r_0^2\cos^2 \th)} \
\ee
and $d\Om_3^2$ is as before. This metric has a horizon at
\be
\label{4-19} u_H^2 = {1 \ov 2} \left( - r_0^2 + \sqrt{r_0^4 + 4
\m^4} \right)\ ,
\ee
and thermodynamic stability restricts $\l \equiv \m / r_0$ to the
range
\be
\label{4-20} \l \gtrsim 0.685 \hbox{ (CE)}\ ,\qq \l \gtrsim 0.93
\hbox{ (GCE)}\  ,
\ee
for the canonical and grand-canonical ensembles respectively. For
the stability analysis, the ``transverse'' and ``angular''
coordinates are as before, but now it is $\phi_1$ that couples to
the longitudinal coordinate $y$. Again, the allowed trajectories
have $\th_0=0$ or $\th_0=\pi/2$, and only the latter case will be
considered here.

\no Using the same rescalings as before, we find that the
expressions for the length and energy read
\be
\label{4-21} L(u_0,\l) = 2 \sqrt{u_0^4-\l^4} \int_{u_0}^\infty {du
\ov \sqrt{(u^4-u_0^4)(u^4+u^2-\l^4)}}\ ,
\ee
and
\be
\label{4-22} E(u_0,\l) = \int_{u_0}^\infty du {u^4-\l^4 \ov
\sqrt{(u^4-u_0^4)(u^4+u^2-\l^4)}}\ - \int_{u_H}^\infty du
\sqrt{{u^4-\l^4 \ov u^4+u^2-\l^4}}\ ,
\ee
which again cannot be evaluated in closed form. Their behavior is
as in Fig.~1c.
\begin{figure}[!t]
\begin{center}
\begin{tabular}{c}
\includegraphics[height=6cm] {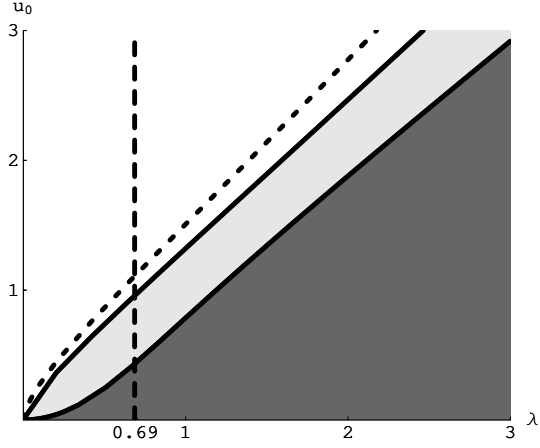}
\end{tabular}
\end{center}
\vskip -.5 cm \caption{Stability diagram in the $(\l,u_0)$ plane
for the spinning D3-brane with one angular momentum and for the
trajectory with $\th_0=\pi/2$. The light-shaded region corresponds
to instabilities under longitudinal perturbations. The dark-shaded
region corresponds to values of $u_0$ beyond the horizon. The
region to the left of the vertical dashed line corresponds to the
values $\l \lesssim 0.685$ for which the background is
thermodynamically unstable. The dotted line corresponds to the
curve drawn using the approximation described below \eqn{4-23}.}
\label{fig4}
\end{figure}

\no Turning to the stability analysis, the coupled system of the
$(\d y,\d \phi_1)$ fluctuations has instabilities for for $u_0 <
u_{0 \rm c}(\l)$ where $u_{0 \rm c}(\l)$ is the critical point of
$L(u_0,\l)$, obtained by numerically solving Eq. \eqn{3-15a} which
for our case reads
\be
\label{4-23} \int_{u_0}^\infty du { u^8 - 4\l^4 u^4 - 4\l^4 u^2 +
3 \l^8 \ov u^4 \sqrt{(u^4-u_0^4)(u^4+u^2-\l^4)^3}} = 0\ .
\ee
This equation has a solution only if the integrand changes sign at
least once in the interval $u\in [u_0,\infty)$. This leads to the
inequality $u_{H}<u_0<u_{\rm max}(\l)$ where $u_{\rm max}(\l)$ is
the largest root of the numerator for which we will not present an
explicit expression due to its complexity. As before, saturation
of the upper bound gives a curve defined by $u_{0 \rm
c}(\l)=u_{\rm max}(\l)$ which gives a quite good approximation to
the exact numerical result, as one can see by inspection of
Fig.~4.

\no Finally, for the angular fluctuations, the Schr\"odinger
potential reads
\be
\label{4-24} V_\th = 1\ ,
\ee
implying that no angular instabilities occur. The results of the
stability analysis are summarized in the diagram of Fig.~4, which
is qualitatively similar to Fig.~3 apart from the absence of
angular instabilities.

\subsection{Marginally-deformed D3-branes}

We finally consider the Lunin--Maldacena deformations \cite{LM} of
D3-brane solutions, dual to the Leigh--Strassler \cite{LS}
marginal deformations of $\cN=4$ SYM which break supersymmetry
down to $\cN=1$. These backgrounds are characterized by the
complex parameter $\b = \g + \tau \s$ (where $\tau$ is the Type
IIB axion-dilaton and $\g$ and $\s$ are real parameters) which, on
the gauge-theory side, represents the complex phase entering the
Leigh--Strassler superpotential (where $\tau$ is the complexified
gauge coupling). In particular, we will consider marginal
deformations of the conformal ${\rm AdS}_5 \times {\rm S}^5$
background \cite{LM} and of the multicenter backgrounds corresponding to
D3-branes distributed on a sphere and on a disc \cite{asz2}.

\subsubsection{The conformal case}

Starting from the deformation of the conformal ${\rm AdS}_5 \times
{\rm S}^5$ background, we rescale the deformation parameters as
$(\b,\g,\s) \to {2 \ov R^2} (\b,\g,g_{\rm s}\s)$ and we write the
deformed metric as
\be
\label{4-25} ds^2 = \cH^{1/2} \left[ {u^2\ov R^2} \left( -  dt^2 +
dx^2 + dy^2 + dz^2 \right) + R^2 \left(  {du^2 \ov u^2} +
d\Omega_{5,\b}^2 \right) \right]\ ,
\ee
where $d\Omega_{5,\b}^2$ is the metric on the deformed ${\rm
S}^5$, given by
\ba
\label{4-26} d\Omega_{5,\b}^2 &=& d \th^2 + \cG \sin^2 \th d
\phi_1^2 + \cos^2 \th [ d \psi^2  + \cG ( \sin^2 \psi d\phi_2^2 +
\cos^2 \psi d \phi_3^2 ) ] \nonumber\\
&+& \cG |\b|^2 \cos^4 \th \sin^2 \th \sin^2 2 \psi
\left(d\phi_1+d\phi_2+d\phi_3 \right)^2\ ,
\ea
and the functions $\cG$ and $\cH$ are given by
\ba
\label{4-27} \cG^{-1} &=& 1 + 4 |\b|^2 \cos^2 \th (\sin^2 \th +
\cos^2 \th \cos^2 \psi \sin^2 \psi)\ , \nonumber\\ \cH &=& 1 +
4\s^2 \cos^2 \th (\sin^2 \th + \cos^2 \th \cos^2 \psi \sin^2
\psi)\ .
\ea
The solution also includes a nonzero $B$--field which, in the case
when the deformation parameter $\g$ vanishes, is equal to
\be
\label{abcd} B_2 = {\s \ov 2}
\cos^4\th\sin{2\psi}(d\phi_1+d\phi_2+d\phi_3)\wedge d\psi\ ,
\ee
and hence is of the form \eqn{2-1a}. For the stability analysis
for this and all subsequent metrics, the ``transverse''
coordinates are $(x,z,\phi_{1,2,3})$, with all of them being
decoupled from the longitudinal coordinate $y$ but with the
$\phi_i$ coupled to each other, while the ``angular'' coordinates
are $(\th,\psi)$, and are decoupled from each other. The
restriction \eqn{2-8} on the angular location $(\th_0,\psi_0)$ of
the string allows only for the trajectories
$(\th_0,\psi_0)=(0,\pi/4)$, $(\th_0,\psi_0)=(0,0 \hbox{ or
}\pi/2)$ (with both choices for $\psi_0$ leading to equivalent
results), $(\th_0,\psi_0)=(\pi/2,{\rm any})$, and
$(\th_0,\psi_0)=(\sin^{-1}(1 / \sqrt{3}),\pi/4)$. All trajectories
but the last will also be valid for the multicenter case with the
branes distributed on the sphere and on a disc. We also note that
for the trajectories with $\th_0=\pi/2$ the variable $\psi$
becomes a completely decoupled ``transverse'' coordinate and hence
its fluctuations are stable. Also, the presence of the nonzero
$B$--field in \eqn{abcd} implies that, for the trajectories with
$(\th_0,\psi_0)=(0,\pi/4)$ and $(\th_0,\psi_0)=(\sin^{-1}(1 /
\sqrt{3}),\pi/4)$, the angular fluctuation $\d \psi$ and the
transverse fluctuations $\d \phi_i$ are actually coupled. However,
the coupled terms are proportional to the eigenvalue $\omega$ and
hence do not affect at all the results of our analysis which is
based on the zero modes. For $(\th_0,\psi_0)=(0,0\hbox{ or }
\pi/2),(\pi/2,\text{any})$, $\d \psi$ and $\d \phi_i$ are
decoupled.

\no For the present case, the potential energy $E$ can be
calculated explicitly as a function of $L$ with the result
\ba
\label{4-28} E(L) = k(\s) \left(- {4\pi^2R^2\ov\Gamma(1/4)^4}{1\ov
L} \right)\ ,
\ea
where $k(\s)$ is an angle- and $\s$--dependent factor, given
explicitly in the examples below, that becomes unity for $\s=0$.
The factor in parentheses is the result of \cite{maldaloop},
giving the standard Coulomb behavior expected by conformal
invariance. We note in passing that this factor is unaffected by
turning on $\g$--deformations, unless the quarks are given a
separation in the internal deformed sphere as well \cite{hsz}.

\no
Regarding stability, the transverse/longitudinal fluctuations
are obviously stable, while the Schr\"odinger potentials for the
angular fluctuations all have the form
\be
\label{4-29} V_{\theta,\psi}= a_{\theta,\psi}(\s) u^2\ .
\ee
where $a_{\th,\psi}(\s)$ are angle- and $\s$--dependent factors
that can be read off the formulas of appendix A. The expression
for the Schr\"odinger variable $x$ in terms of $u$ is
\ba
\label{4-30} x={1\ov u}\ {}_2F_1\left({1 \ov 4},{1 \ov 2},{5 \ov
4},{u_0^4 \ov u^4} \right)
\ea
and, accordingly, the value of the endpoint $x_0$ is
\ba
\label{4-31} x_0={\Gamma(1/4)^2\ov 4 \sqrt{2\pi}}{1\ov u_0}\ .
\ea
Instabilities may occur only when one of $a_{\th,\psi}(\s)$ is
negative. Although Eq. \eqn{4-30} cannot be inverted in terms of
$u$ to allow for an analytic solution of the Schr\"odinger
problem, we may obtain a lower bound for $\s$ above which
instabilities definitely occur. We first note that the
infinite-well approximation is not valid here, as the potential
satisfies Eq. \eqn{3-22} with $a<0$. To obtain our bound, we
consider the limit $u/u_0 \gg 1$, where $x \simeq 1/u$ and
$V_{\th,\psi} \simeq a_{\th,\psi}(\s) / x^2$. Then, standard
arguments from quantum mechanics \cite{LLQM} show that for,
$a_{\th,\psi}(\s) < - {1 \ov 4}$, this potential supports an
infinite tower of negative-energy states and the solution becomes
unstable for all $u_0$. This gives the desired bound on $\s$. The
results for the allowed trajectories are as follows:

\no
$\bullet$
$(\th_0,\psi_0)=(0,\pi/4)$. For this case we have
\be
\label{4-32} k(\s) = \sqrt{1+\s^2}\ ,\qq a_\th(\s) = {2\s^2 \ov
1+\s^2}\ ,\qq a_\psi(\s) = - {4\s^2 \ov 1+\s^2}\ .
\ee
Instabilities occur only from the $\d \psi$ fluctuations for $\s >
1 / \sqrt{15} \simeq 0.258$. Since in the UV all marginally
deformed backgrounds approach \eqn{4-25}, \eqn{4-26}, the above
bound is universal as long as the corresponding trajectory remains
valid. This is indeed the case for the sphere and disc brane
distributions we consider.

\no $\bullet$ $(\theta_0,\psi_0)=(0,0 \hbox{ or } \pi/2)$. Here,
we have
\be
\label{4-33} k(\s) = 1\ ,\qq a_\th(\s) = a_\psi(\s) = 4\s^2\ ,
\ee
and no angular instabilities occur.

\no $\bullet$ $(\th_0,\psi_0)=(\pi/2,{\rm any})$. Here, we have
\be
\label{4-34} k(\s) = 1\ ,\qq a_\th(\s) = 4 \s^2\ ,
\ee
and again no angular instabilities occur.

\no $\bullet$ $(\th_0,\psi_0)=(\sin^{-1}(1/\sqrt{3}),\pi/4)$.
Here, we have
\be
\label{4-35} k(\s) = \sqrt{1+{4\s^2 \ov 3}}\ ,\qq a_\th(\s) =
a_\psi(\s) = - {8 \s^2 \ov 3+4\s^2}\ .
\ee
Instabilities occur from both $\d \th$ and $\d \psi$ fluctuations
for $\s > \sqrt{3/28} \simeq 0.327$. This bound does not survive
in the multicenter cases we consider below since the corresponding
trajectory is no longer valid.

\no Note that the existence of un upper bound for the deformation
parameter $\s$ beyond which stability breaks down is reminiscent
of an analogous fact for giant gravitons on the pp-wave limit of
the deformed background \eqn{4-25}, \eqn{4-26}, which exist only
for values of $\s< 1/\sqrt{12}$. These giant graviton solutions
can be thought of as integrated perturbations around their
infinitesimal size. This implies that every order in the small
size perturbative expansion of the giant graviton solution is
unstable if the above bound is not respected. The different upper
value for $\s$ in this case should be attributed to the fact that
the probes in the giant graviton case refer to D3-branes and not
to strings.

\subsubsection{The sphere}

We next consider the deformation of the background corresponding
to D3-branes distributed on a sphere of radius $r_0$. Switching to
dimensionless units by using Eq. \eqn{4-11}, we write the deformed
metric as \cite{asz2}
\ba
\label{4-36} &&\!\!\!\!\!\!\!\!\!\!\!\!\!\! ds^2 = \cH^{1/2}
\Biggl\{ H^{-1/2} \left( - dt^2 + dx^2 + dy^2 + dz^2 \right) +
H^{1/2} {u^2 - r_0^2 \cos^2 \th \ov u^2 - r_0^2} du^2
\nonumber\\
&&\!\!\!\!\!\!\!\!\!\!\!\!\!\!\ \qq\qq\quad + \:\: H^{1/2}
\left[(u^2 - r_0^2 \cos^2 \th) d \th^2 + \cG (u^2 - r_0^2) \sin^2
\th d \phi_1^2 \right]
\nonumber\\
&&\!\!\!\!\!\!\!\!\!\!\!\!\!\!\ \qq\qq\quad + \:\: H^{1/2} u^2
\cos^2 \th \left[ d \psi^2 + \cG ( \sin^2 \psi d\phi_2^2 + \cos^2
\psi d \phi_3^2 ) \right]
\nonumber\\
&&\!\!\!\!\!\!\!\!\!\!\!\!\!\!\ \qq\qq\quad + \:\: H^{1/2} {|\b|^2
\cG u^2 (u^2-r_0^2) \cos^4 \th \sin^2 \th \sin^2 2 \psi \ov u^2 -
r_0^2 \cos^2 \th} \left(d\phi_1+d\phi_2+d\phi_3 \right)^2
\Biggr\}\ ,
\ea
with
\ba
\label{4-37} \cG^{-1} &=& 1 + 4|\b|^2 \cos^2 \th\ { ( u^2 - r_0^2
) \sin^2 \th + u^2 \cos^2 \th \cos^2 \psi \sin^2 \psi  \ov u^2 -
r_0^2 \cos^2 \th}\ , \nonumber\\ \cH &=& 1 + 4\s^2 \cos^2 \th\ { (
u^2 - r_0^2 ) \sin^2 \th + u^2 \cos^2 \th \cos^2 \psi \sin^2 \psi
\ov u^2 - r_0^2 \cos^2 \th}\ ,
\ea
and $H$ is given by \eqn{4-8}. We also note that here there is a
nonzero $B$--field which for $\g=0$ is proportional to the
expression given in \eqn{abcd}. Now, the restriction \eqn{2-8} on
the $(\th_0,\psi_0)$ of the string allows only for the first three
trajectories considered earlier, namely
$(\th_0,\psi_0)=(0,\pi/4)$, $(\th_0,\psi_0)=(0,0 \hbox{ or
}\pi/2)$ and $(\th_0,\psi_0)=(\pi/2,{\rm any})$.

\no We next examine the potentials arising in each case in turn:

\no $\bullet$ $(\th_0,\psi_0)=(0,\pi/4)$. The length and potential
energy read (see sec. 6.2 of \cite{asz2})
\be
\label{4-38} L(u_0,\s) = { 2 u_0 \ov \sqrt{(u_0^2-{1 \ov
1+\s^2})(2 u_0^2-{1 \ov 1+\s^2})} } \left[ \elPi(a^2,k) - \elK(k)
\right]\ ,
\ee
and
\be
\label{4-39} E(u_0,\s) = \sqrt{1+\s^2} \left\{ \sqrt{2 u_0^2-{1
\ov 1+\s^2}} \left[ a^2 \elK(k) - \elE(k) \right] + \elE(c)-{\s^2
\ov 1+\s^2}\elK(c) \right\}\ ,
\ee
where
\be
\label{4-40} k = \sqrt{{u_0^2 + {\s^2 \ov 1+\s^2} \ov 2 u_0^2 - {1
\ov 1+\s^2}}}\ ,\qq a = \sqrt{{u_0^2 - {1 \ov 1+\s^2} \ov 2 u_0^2
- {1 \ov 1+\s^2}}}\ , \qq c={1 \ov \sqrt{1 + \s ^2}}\ .
\ee
Note that these results are independent of the deformation
parameter $\g$. The reason is that, as remarked also in the
conformal case, we have not separated the quarks in the internal
deformed-sphere space. Wilson-loop potentials where such a
separation was considered, but with vanishing $\s$--deformation,
were computed in \cite{hsz}.

\no For $\s=0$, the behavior of the length and energy is the same
as for the undeformed case (Fig.~1c). As $\s$ is turned on, the
length and energy curves are reminiscent of van der Waals
isotherms for a statistical system with $u_0$, $L$ and $E$
corresponding to volume, pressure and Gibbs potential respectively
(see e.g. \cite{Callen}). In particular, there exists a critical
value of $\s$, given by $\s_{\rm cr}^{(y)}\simeq 0.209$ (found
analytically in \cite{asz2}), below which the system behaves like
the statistical system at $T<T_{\rm cr}$ (Fig.~1d) and above which
the system behaves like the statistical system at $T>T_{\rm cr}$
(Fig.~1b). For nonzero $\s$, we have the usual Coulombic behavior
for large $u_0$ (small $L$), while in the opposite limit, $u_0 \to
1$, (large $L$), the asymptotics of \eqn{4-38} and \eqn{4-39} lead
to the linear potential
\be
\label{4-41} E \simeq {\s \ov 2} L\ .
\ee
To examine the significance of these results, it is crucial to
examine the stability of the corresponding string configurations.

\no Our general results imply that the solution is stable under
transverse perturbations. For the longitudinal fluctuations, the
fact that $L(u_0)$ has two extrema $u_{0 \rm c}^{(l1)}(\s)$ and
$u_{0 \rm c}^{(l2)}(\s)$ for $0<\s<\s^{(y)}_{\rm cr}$ and no
extrema for $\s > \s^{(y)}_{\rm cr}$ leads us to expect
longitudinal instabilities in the region $u_{0 \rm c}^{(l1)}(\s) <
u_0 < u_{0 \rm c}^{(l2)}(\s)$ of $u_0$ (cf. Fig.~1d) in the first
case, and no instabilities (cf. Fig.~1b) in the second
case.\footnote{For $\s>1/4$ the derivative term in \eqn{3-15a} has
definite sign, so \eqn{3-15a} has no solution for any value of
$u_0$.} To verify that the lowest eigenvalue $\omega_0^2$ does
indeed change sign at these values, we performed a numerical
analysis whose results are shown in Fig.~5b and indeed reproduce
the expected behavior.\footnote{For this case, we can use a
Schr\"odinger description for the longitudinal fluctuations and
apply our infinite-well approximation. This does indeed lead to
two critical values of $u_0$ for $0<\s<\s^{(y)}_{\rm cr}$.}
Finally, for the angular fluctuations, the relevant Schr\"odinger
potentials are given by the complicated expressions in Eqs.
\eqn{B-12} of appendix A. Starting from $V_\th$, we find that for
$\s<0.71$ it is positive while for $\s>0.71$ it develops a
negative part which means that it can in principle support a bound
state of negative energy. The critical value $\s^{(\th)}_{\rm cr}$
above which such states can exist is determined in the
infinite-well approximation to be $\s^{(\th)}_{\rm cr} \simeq
0.78$, which is quite close to the true value $\s^{(\th)}_{\rm cr}
\simeq 0.805$ determined by numerical analysis. As $\s$ increases
beyond this value, the corresponding critical value $u_{0\rm
c}^{(\th)}$ approaches the value of 1.097 as $\s$ becomes very
large. Turning to the $\d\psi$ fluctuations, the same reasoning
used in the deformation of the conformal background for the same
trajectory carries over to the present case, implying that there
exists a critical value $\s^{(\psi)}_{\rm cr} = 1/\sqrt{15}$
(within our numerical limitations) above which there occur
instabilities for all values of $u_0$. This value is presumably
inherited by the conformal limit of the potential as analyzed in
subsection 4.3.1. The above results are summarized in the
stability diagram of Fig.~5.
\begin{figure}[!t]
\begin{center}
\begin{tabular}{ccc}
 \includegraphics[height=4.7cm]{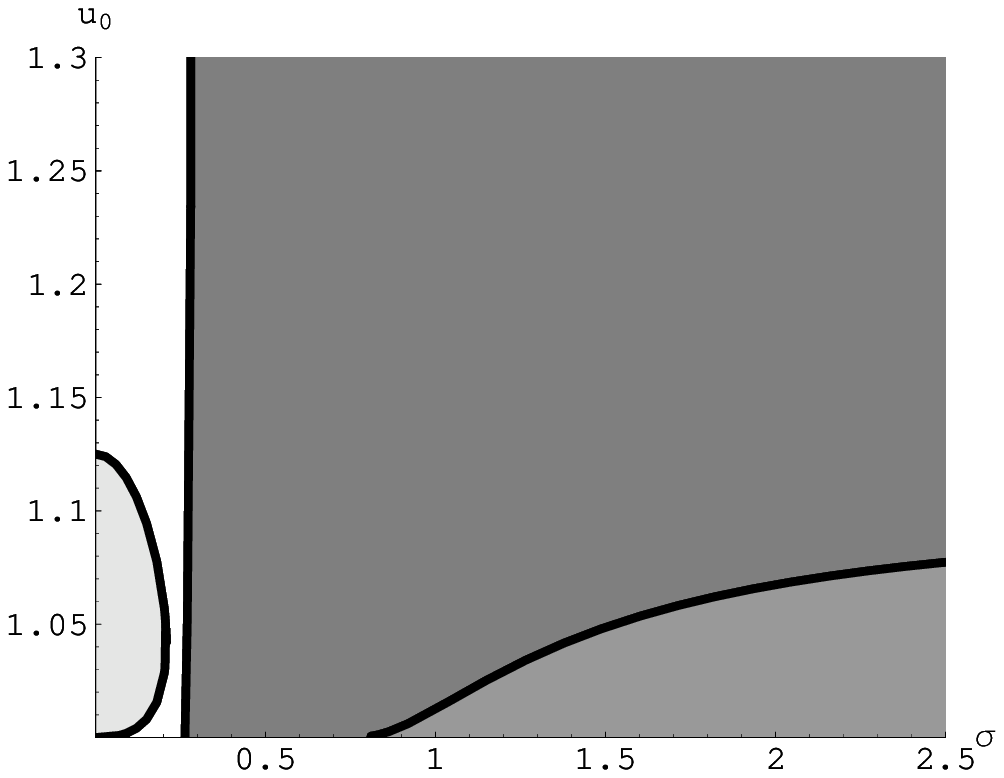}
&\includegraphics[height=4.7cm]{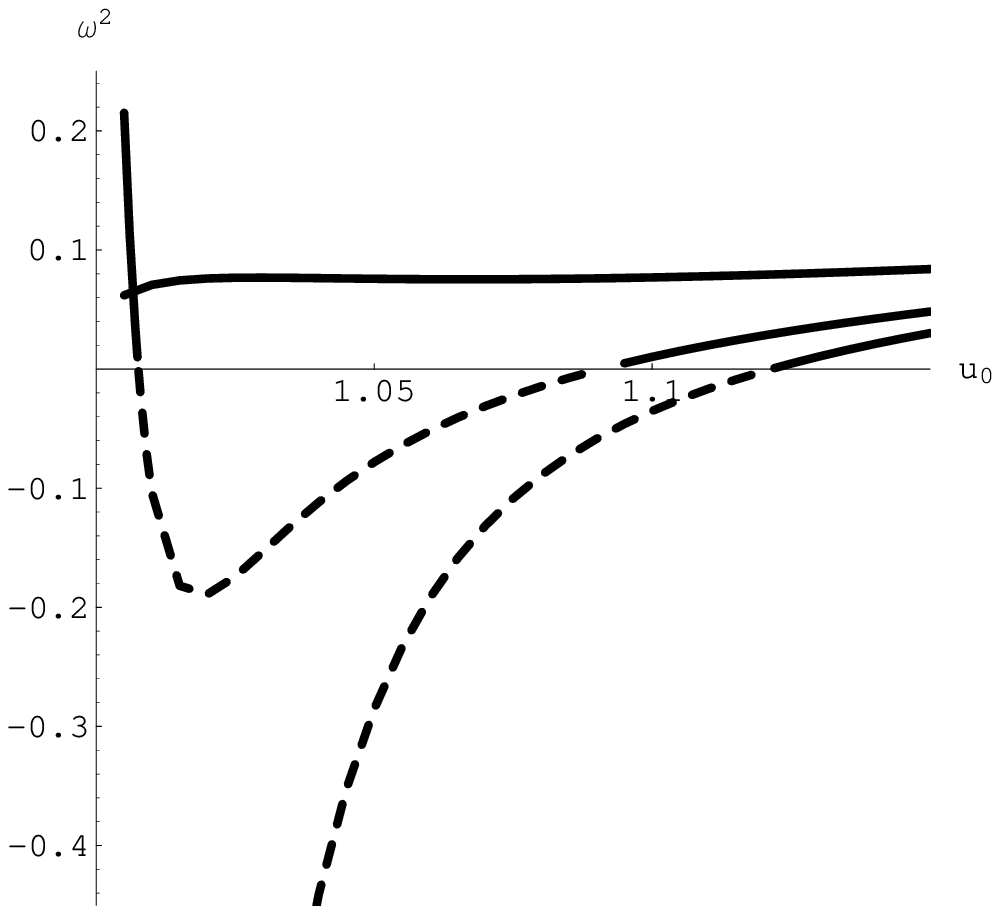}
&\includegraphics[height=4.7cm]{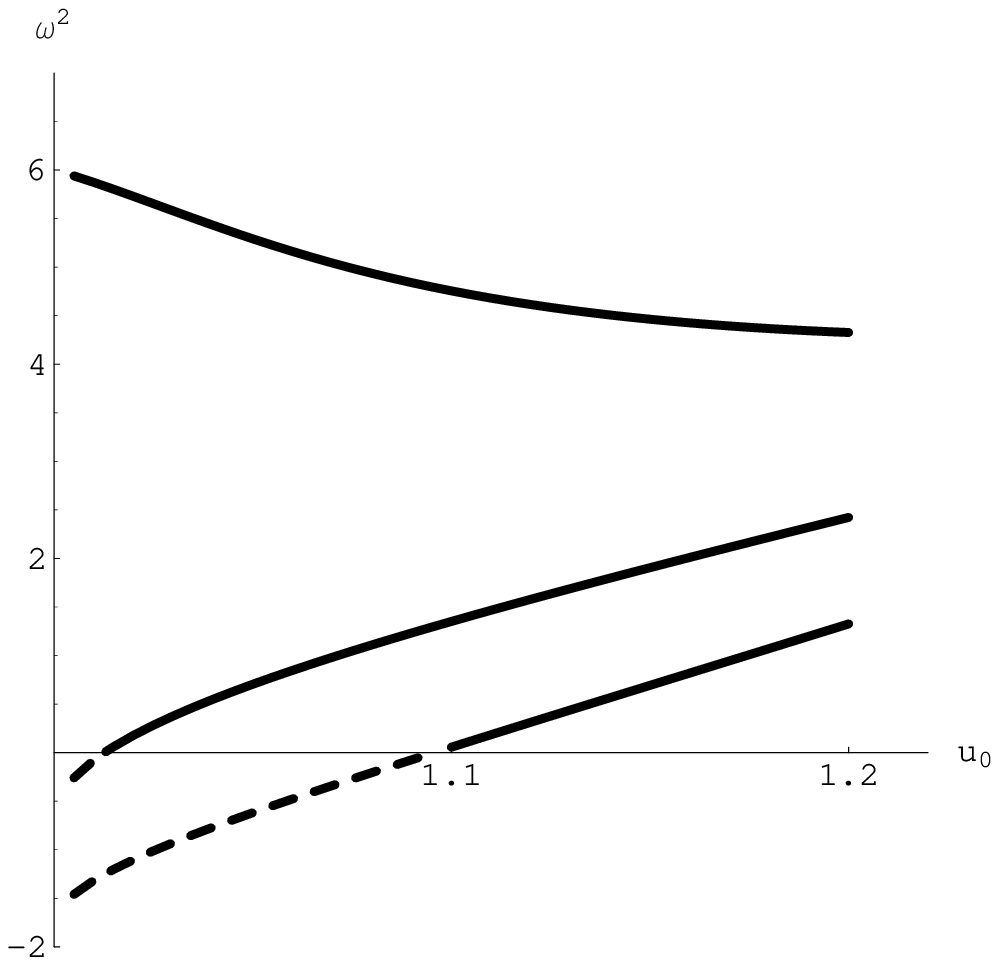}\\
(a) & (b) & (c)
\end{tabular}
\end{center}
\vskip -.5 cm \caption{(a) Stability diagram in the $(\s,u_0)$
plane for the deformed sphere and the trajectory with
$(\th_0,\psi_0)=(0,\pi/4)$. The three shaded regions correspond to
instabilities due to longitudinal, $\d \th$ and $\d \psi$
fluctuations (lightest to darkest). The origin $(0,1)$ actually
belongs to the stable part of the diagram. Specifically, it is
stable on its own right, but nowhere in its vicinity (for an
analogous explicit example see sec. 5.4 of \cite{stability1}). (b)
Evolution of the lowest eigenvalue of the longitudinal
fluctuations with $u_0$, plotted for $\s=0$, $0.15 < \s^{(y)}_{\rm
cr}$ and $0.5 > \s^{(y)}_{\rm cr}$ (bottom to top). (c) Same for
the angular $\d \th$ fluctuations, for $\s=0<\s^{(\th)}_{\rm cr}$,
$1 > \s^{(\th)}_{\rm cr}$ and $10 \gg \s^{(\th)}_{\rm cr}$ (top to
bottom).} \label{fig5}
\end{figure}

\no We also note that, in the limit $\s \gg 1$, the potential for
the $\d\th$ fluctuations reads
\be
\label{4-42} V_\th=2(u^2-2)+{\cal O}(\s^{-2})\ ,
\ee
and $x$ can be explicitly determined in terms of $u$ by
\eqn{4-51}. In this limit, the problem can be solved exactly (see
appendix B), leading to the critical value $u_{0\rm
c}^{(\th)}\simeq 1.09737$, by mapping it to the well studied
Lam\'e equation.

\no From the above we conclude that for a finite narrow range of
values of the deformation parameter, namely for $0.209 \simeq
\s^{(y)}_{\rm cr} < \s  < \s^{(\psi)}_{\rm cr} \simeq 0.258$ the
linear confining behavior for large distances is stable, after
which it is wiped out first by the instability of the $\d\psi$
fluctuations and subsequently by the $\d \th$ ones. Moreover, even
for smaller values of $\s$, namely for $0< \s< \s^{(y)}_{\rm
cr}\simeq 0.209 $, the confining behavior of potential is stable
as we have discussed already.

\no $\bullet$ $(\th_0=0,\psi_0=0 \hbox{ or } \pi/2)$. The length
and potential energy are the same as in the corresponding
undeformed case, namely they are given by \cite{bs}
\be
\label{4-43} L(u_0) = {2 u_0 k^{\prime} \ov u_0^2-1} \left[ \elPi
(k^{\prime 2},k) - \elK(k) \right]\ ,
\ee
and
\be
\label{4-44} E(u_0) = \sqrt{2u_0^2-1} \left[ k^{\prime 2} \elK (k) -
\elE(k) \right] + 1\ ,
\ee
where
\be
\label{4-45} k={u_0 \ov \sqrt{2u_0^2-1}}\ ,\qq
k^{\prime}=\sqrt{1-k^2}\ .
\ee
Their behavior is as in Fig.~1c.

\no By our general results, there exist longitudinal instabilities
occurring for $u_0$ below the critical value for the undeformed
case, namely $u_{0 \rm c}\simeq 1.125$. For the angular $\d\th$
and $\d\psi$ fluctuations, the Schr\"odinger potentials are given
by \eqn{B-13} of appendix A and differ by the ones in the
undeformed case by the positive-definite term $4 \s^2 u^2$, while
the expression for the Schr\"odinger variable $x$ in terms of $u$
is the same as in the undeformed case. The above results imply
that, since the $\d\th$ and $\d\psi$ fluctuations are stable in
the undeformed case, they are stable in the present case as well.

\no $\bullet$ $(\th_0,\psi_0)=(\pi/2,{\rm any})$. In this case,
the length and potential energy are again the same as in the
undeformed case and are given by \cite{bs}
\be
\label{4-46} L(u_0) = {\sqrt{2} \ov u_0} \left[ \elPi \left( \ha,k
\right) -\elK(k) \right]\ ,
\ee
and
\be
\label{4-47} E(u_0) = {u_0 \ov \sqrt{2}} \left[ \elK(k)- 2 \elE(k)
\right]\ .
\ee
where
\be
\label{4-48} k = \sqrt{{u_0^2+1 \ov 2 u_0^2}}\ ,\qq
k^{\prime}=\sqrt{1-k^2}\ .
\ee
The behavior is as shown in Fig.~1b. For $u_0 \gg 1$ (small $L$),
the behavior is Coulombic, whereas in the opposite limit, $u_0 \to
1$ (large $L$), the asymptotics of \eqn{4-46} and \eqn{4-47} lead
to the linear potential
\be
\label{4-49} E \simeq  {L \ov 2}\ ,\qq {\rm for} \quad L \gg 1 \ .
\ee
In the undeformed case, the appearance of a linear confining
potential is quite unexpected from the gauge-theory side and the
paradox is resolved by the stability analysis of \cite{stability1}
which indicates that the configurations where this behavior
appears are unstable under angular perturbations. However, for the
deformed case, where $\cN=4$ supersymmetry is broken to $\cN=1$,
the appearance of a confining potential is expected. Remarkably,
this is confirmed by the stability analysis that follows:

\no We first note that the transverse and longitudinal
fluctuations are manifestly stable, the latter fact following from
the absence of extrema of $L(u_0)$. For the angular fluctuations,
the Schr\"odinger potential reads
\be
\label{4-50} V_\th = -1 + 4 \s^2(u^2-1)\ ,
\ee
whence we see that the effect of the deformation is to modify
$V_\th$ (which, for the undeformed case equals $-1$) by a
positive-semidefinite term. It is quite intuitive that the
addition of this term will tend to stabilize the angular
fluctuations. To see how this occurs, we note that in the present
case, the Schr\"odinger variable $x$ can be explicitly determined
in terms of $u$ as
\be
\label{4-51} x = \sqrt{2 k^2-1\ov 2}\ {\bf F}(\nu,k)\ , \qq
\nu=\sin^{-1}\sqrt{2u_0^2\ov u^2+u_0^2}\ ,
\ee
where $\elF(\n,k)$ is
the incomplete elliptic integral of the first kind and $k$ is the modulus defined in \eqn{4-48}.
Likewise, the endpoint $x_0$ is determined in terms of $u_0$ by
\be
\label{4-52} x_0=x(u_0)=\sqrt{2 k^2-1\ov 2}\ {\bf K}(k)\ .
\ee
Then, as shown in detail in appendix B, the equation for the
angular fluctuations takes the form \eqn{A-1} with $A$ and $h$
given by the second of \eqn{A-3}. For general values of $\s$, this
equation can be solved only numerically and the results are as
shown in Fig.~6. We verify that the critical point $u_{0 \rm c}$
below which instabilities occur is a decreasing function of $\s$
which implies that the region where confinement occurs is
gradually stabilized as one increases $\s$.

\no We finally note that for the special values $4\s^2=n(n+1)$,
where $n$ is a positive integer, \eqn{A-1} reduces to the Lam\'e
equation and can be solved exactly, as done in appendix B for the
particular case $n=1$. The resulting value of $u_{0 \rm c}$, given
in the second of \eqn{A-13}, agrees with our numerical results.
\begin{figure}[!t]
\begin{center}
\begin{tabular}{cc}
\includegraphics[height=6cm]{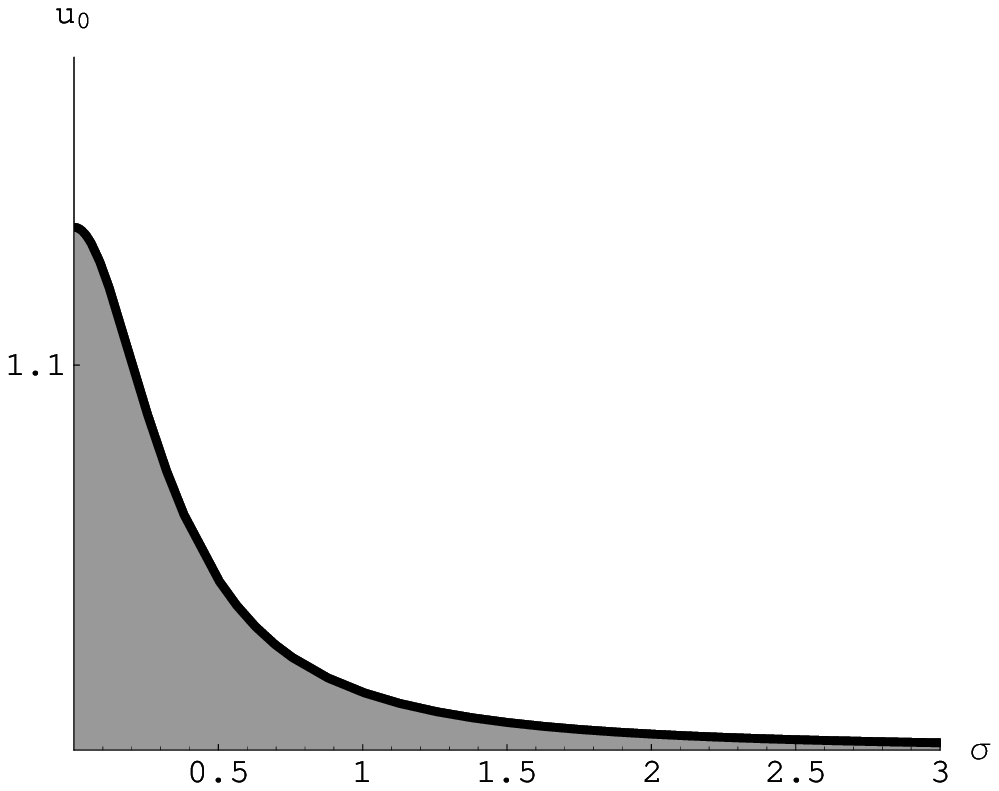}
&\includegraphics[height=6cm]{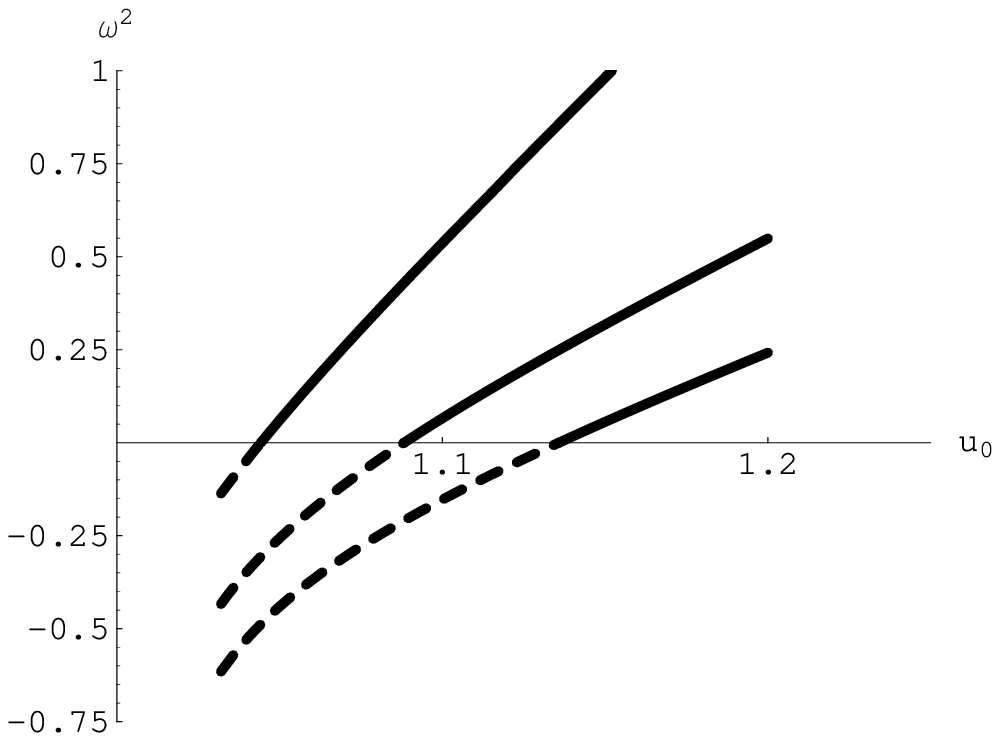}\\
(a) & (b)
\end{tabular}
\end{center}
\vskip -.5 cm \caption{(a) Stability diagram in the $(\s,u_0)$
plane for the deformed sphere and the trajectory with
$(\th_0,\psi_0)=(\pi/2,{\rm any})$. The shaded region corresponds
to instabilities under angular perturbations. (b) Evolution of the
lowest eigenvalue of the angular fluctuations with $u_0$, plotted
for $\s=0$, $0.25$ and $0.5$ (bottom to top).} \label{fig6}
\end{figure}

\subsubsection{The disc}

We finally consider the deformation of the background
corresponding to D3-branes distributed on a disc of radius $r_0$,
which is obtained from the deformed sphere background by the
analytic continuation $r_0^2 \to -r_0^2$. The possible
trajectories are the same as for the sphere. We examine them in
turn:

\no $\bullet$ $(\th_0,\psi_0)=(0,\pi/4)$. The length and potential
energy are given by (see sec. 6.2 of \cite{asz2})
\be
\label{4-53} L(u_0,\s) = { 2 u_0 \ov \sqrt{(u_0^2+{1 \ov
1+\s^2})(2 u_0^2+{1 \ov 1+\s^2})} } \left[ \elPi(a^2,k) - \elK(k)
\right]\ ,
\ee
and
\be
\label{4-54} E(u_0,\s) = \sqrt{1+\s^2} \left\{  \sqrt{2 u_0^2+{1
\ov 1+\s^2}} \left[ a^2 \elK(k) - \elE(k) \right] + \elE(c)-{1 \ov
1+\s^2}\elK(c) \right\}\ ,
\ee
where
\be
\label{4-55} k = \sqrt{{u_0^2 - {\s^2 \ov 1+\s^2} \ov 2 u_0^2 + {1
\ov 1+\s^2}}}\ ,\qq a = \sqrt{{u_0^2 + {1 \ov 1+\s^2} \ov 2 u_0^2
+ {1 \ov 1+\s^2}}}\ , \qq c={\s \ov \sqrt{1 + \s ^2}}\ .
\ee
For $u_0 \gg 1$ (small $L$) we recover the standard Coulombic
behavior enhanced by the factor $\sqrt{1+\s^2}$, while for $u_0
\to 0$ the asymptotics lead to the potential
\be
\label{4-56} E \simeq - {(\pi - L)^2 \ov 8 \elE({\rm i} \s)}\ ,
\ee
which shows that there is complete screening at the screening
length
\be
L_{\rm c} = \pi\ ,
\label{lps}
\ee
that is invariant under $\s$--deformations.

\no Regarding stability, the solution is stable under transverse
and longitudinal perturbations. Turning to angular fluctuations,
the relevant Schr\"odinger potentials are given by the complicated
expressions in Eqs. \eqn{B-15} of appendix A. Starting from the
$\d\th$ fluctuations, we use the infinite-well approximation to
find that the bottom of the well is raised with increasing $\s$
and that there is a critical value $\s^{(\th)}_{\rm
cr}>1/\sqrt{15}$ above which no instabilities exist. The existence
of this critical value can be demonstrated by noting that for $\s
\gg 1$, the potential takes the form
\ba
\label{4-57} V_{\th} \simeq 2(u^2+2)+{\cal O}(\s^{-2})\ .
\ea
This potential is positive and, using the techniques of appendix
B, one can then show that perturbation theory holds about $\s \to
\infty$. Thus, no instabilities occur in this limit. For the
$\d\psi$ fluctuations, the same arguments used for the deformed
conformal background for the same trajectory show that there
exists another critical value $\s^{(\psi)}_{\rm cr} = 1/\sqrt{15}$
(again within numerical limitations and as before inherited by the
conformal limit behavior of the solution) above which there occur
instabilities for all values of $u_0$.

\no $\bullet$ $(\th_0,\psi_0)=(0,0 \hbox{ or } \pi/2)$. In this
case we find that the classical solution is the same as in the
undeformed disc at $\th=0$ \cite{bs}
\be
\label{4-58} L(u_0) = {2 u_0 k^{\prime} \ov u_0^2 +1} \left[ \elPi
(k^{\prime 2},k) - \elK(k) \right]
\ee
and
\be
\label{4-59} E(u_0) = \sqrt{2u_0^2+1} \left[ k^{\prime 2} \elK (k) -
\elE(k) \right] \ ,
\ee
where
\be
\label{4-60} k={u_0 \ov \sqrt{2u_0^2+1}}\ , \qq k^{\prime
}=\sqrt{1-k^2}\ ,
\ee
Their behavior is as in Fig.~1a.

\no By our general results, the solution is stable under
transverse and longitudinal perturbations. For the angular $\d\th$
and $\d\psi$ fluctuations, the Schr\"odinger potentials are given
by \eqn{B-16} of appendix A and again differ by the ones in the
undeformed case by $4 \s^2 u^2$, while the expression for the
Schr\"odinger variable $x$ in terms of $u$ is the same. For the
$\d\th$ fluctuations, which have instabilities in the undeformed
case, the extra term stabilizes the fluctuations and there exists
critical value $\s^{(\th)}_{\rm cr}$ above which there are no
instabilities. Finally, the $\d\psi$ fluctuations are stable in
the undeformed case and hence are stable in the present case as
well.

\no $\bullet$ $(\th_0,\psi_0)=(\pi/2,{\rm any})$. The length and
potential energy are given by \cite{bs}
\be
\label{4-61} L(u_0) = {2 u_<^2 \ov \sqrt{u_0^2+u_>^2}}\left[ {\bf
\Pi} \left({u_>^2\ov u_0^2+u_>^2},k\right)-{\bf K}(k)\right ]\ ,
\ee
and
\be
\label{4-62} E(u_0) = {u_0^2\ov \sqrt{u_0^2+u_>^2}} \ {\bf
K}(k) - \sqrt{u_0^2+u_>^2} \ {\bf E}(k) + 1\ ,
\ee
where now
\be
\label{4-63} k = \sqrt{{u_>^2 -u_<^2 \ov u_0^2+u_>^2}}\ , \qq
k^{\prime}=\sqrt{1-k^2}\
\ee
and $u_>$ ($u_<$) denotes the larger (smaller) between $u_0$ and
$1$. Their behavior is as in Fig.~1a and, in particular, we have a
screened Coulomb potential with screening length
\be
\label{4-64} L_{\rm c} = {\pi \ov 2}\ .
\ee

\no By familiar arguments, the solution is stable under transverse
and longitudinal perturbations, whose behavior is insensitive to
the deformation. Regarding the angular fluctuations, the
Schr\"odinger potential (equal to $1$ in the undeformed case) now
reads
\ba
\label{4-65} V_\th = 1 + 4 \s^2(u^2+1)\ ,
\ea
and hence the solution is still stable under angular
perturbations.

\section{Discussion and concluding remarks}

\no In this paper, we completed the analysis of our earlier work
\cite{stability1} by developing a formalism for studying the
perturbative stability of strings dual to quark-antiquark pairs in
a wide class of backgrounds with non-diagonal metrics and possibly
a $B$--field turned on. In particular, we derived a set of results
by means of which the regions where instabilities occur can be
determined by studying the zero modes of the differential
equations governing the fluctuations. The simplifying fact making
this extension possible is that, although the various fluctuations
are generally coupled, their zero modes actually decouple so that
the problem can be treated in an analytic manner similar to the
diagonal case.

\no The methods developed here were applied to strings in boosted,
spinning and marginally-deformed D3-brane backgrounds, dual to
quarkonium states. For the case of boosted D3-branes, we easily
recovered the result that instabilities appear for the
energetically unfavored ``long'' strings stretching further into
the radial direction. For the case of spinning branes, we found
regions of instabilities due to both longitudinal and angular
fluctuations. Finally, for the case of marginally-deformed
D3-branes, we found that angular fluctuations tend to completely
destabilize the classical solutions for large enough values of the
deformation parameter $\s$, except for some special cases where
the classical solutions are unaffected by the deformation and for
which the deformation parameter $\s$ may actually have a
stabilizing effect. In particular, we found parametric regions
giving rise to a linear confining potential for which the dual
string configurations are stable.

\no As the methods developed here are completely general, we
expect that they are directly applicable to other backgrounds and,
in particular to other gravity duals of gauge theories with
reduced or no supersymmetry. As such backgrounds typically involve
a non-trivial dependence on certain angular coordinates, it would
be particularly interesting to determine the parametric regions
for which quarkonium string configurations in these backgrounds
are stable under angular perturbations. Extending our formalism to
fluctuations of general brane probes should be also possible.

\vskip 0.5cm

\centerline{ \bf Acknowledgments}

\no K.~Sfetsos and K.~Siampos acknowledge support provided through
the European Community's program ``Constituents, Fundamental
Forces and Symmetries of the Universe'' with contract
MRTN-CT-2004-005104, the INTAS contract 03-51-6346 ``Strings,
branes and higher-spin gauge fields'', the Greek Ministry of
Education programs $\rm \P Y\Th A\G OPA\S$ with contract 89194.
K.~Siampos also acknowledges support provided by the Greek State
Scholarship Foundation (IKY).

\appendix

\section{Angular Schr\"odinger potentials}

Here we collect, for reference purposes, the Schr\"odinger
potentials for angular fluctuations for all metrics and
trajectories considered in the present paper as well as in
\cite{stability1}.

\subsubsection*{Multicenter D3-branes}

The only ``angular'' coordinate is $\th$. The Schr\"odinger
potentials for the various trajectories are given below.

\no $\bullet$ Sphere, $\th_0=0$:
\be
\label{B-1} V_\th(u;u_0) = {2 u^6 - u^4 + [6 u_0^2(u_0^2-1)-1] u^2
- 3 u_0^2(u_0^2-1) \ov 4 u^2(u^2-1)^2}\ .
\ee
\no $\bullet$ Sphere, $\th_0=\pi/2$:
\be
\label{B-2} V_\th(u;u_0) = -1\ .
\ee
\no $\bullet$ Disc, $\th_0=0$:
\be
\label{B-3} V_\th(u;u_0) = - {2 u^6 + u^4 + [6 u_0^2(u_0^2+1)-1]
u^2 + 3 u_0^2(u_0^2+1) \ov 4 u^2(u^2+1)^2}\ .
\ee
\no $\bullet$ Disc, $\th_0=\pi/2$.
\be
\label{B-4} V_\th(u;u_0) = 1\ .
\ee

\subsubsection*{Spinning D3-branes}

The only ``angular'' coordinate is $\th$ as before. The
Schr\"odinger potentials are given below.

\no $\bullet$ Two equal angular momenta, $\th_0=0$:
\ba
\label{B-5} \!\!\!\!\!\!\!\!V_\th(u;u_0,\l) &=& {1 \ov 4
u^6(u^2-1)^2} \Bigl\{ 2
u^{10} - u^8 + \{6 [ u_0^2(u_0^2-1) + \l^4] -1 \} u^6 \nonumber\\
&-& [3 u_0^2(u_0^2-1) + 11 \l^4] u^4 - 5 [2u_0^2(u_0^2-1)-1] \l^4
u^2 \nonumber\\
&+& 7 u_0^2(u_0^2-1) \l^4 \Bigr\}\ .
\ea
\no $\bullet$ Two equal angular momenta, $\th_0=\pi/2$:
\be
\label{B-6} V_\th(u;u_0,\l) = -1\ .
\ee
\no $\bullet$ One angular momentum, $\th_0=0$:
\ba
\label{B-7} \!\!\!\!\!\!\!\!V_\th(u;u_0,\l) &=& - {1 \ov 4
u^4(u^2+1)^2} \Bigl\{
2 u^{10} + 3 u^8 + \{6 [ u_0^2(u_0^2+1) + \l^4] -1 \} u^6 \nonumber\\
&+& \{ 9 [ u_0^2(u_0^2+1) + \l^4] -1 \} u^4 +
[(3-10\l^4)u_0^2(u_0^2+1)+3\l^4] u^2 \nonumber\\
&-& 5 u_0^2(u_0^2+1) \l^4 \Bigr\}\ .
\ea
\no $\bullet$ One angular momentum, $\th_0=\pi/2$:
\be
\label{B-8} V_\th(u;u_0,\l) = 1\ .
\ee

\subsubsection*{Deformed D3-branes}

Now, the ``angular'' coordinates are $\th$ and $\psi$, with $\psi$
becoming irrelevant for the trajectories with $\th_0=\pi/2$. The
Schr\"odinger potentials are as follows.

\no $\bullet$ Conformal, $(\th_0,\psi_0)=(0,\pi/4)$:
\be
\label{B-9} V_\th(u;u_0,\s) = {2\s^2 \ov 1+\s^2}u^2\ ,\qq
V_\psi(u;u_0,\s) = - {4\s^2 \ov 1+\s^2}u^2\ .
\ee
\no $\bullet$ Conformal, $(\theta_0,\psi_0)=(0,0 \hbox{ or }
\pi/2)$:
\be
\label{B-10} V_\theta(u;u_0,\s)=V_\psi(u;u_0,\s)=4\s^2 u^2\ .
\ee
\no $\bullet$ Conformal, $(\th_0,\psi_0)=(\pi/2,{\rm any})$:
\be
\label{B-11} V_\th(u;u_0,\s) = 4 \s^2 u^2\ .
\ee
\no $\bullet$ Conformal,
$(\th_0,\psi_0)=(\sin^{-1}(1/\sqrt{3}),\pi/4)$:
\be
\label{B-11a} V_\th(u;u_0,\s) = V_\psi(u;u_0,\s) = - {8 \s^2 \ov
3+4\s^2} u^2\ .
\ee
\no $\bullet$ Sphere, $(\th_0,\psi_0)=(0,\pi/4)$:
\ba
\label{B-12} V_\th(u;u_0,\s) &=& {1 \ov 4u^2[(1+\s^2)u^2-1]^3}
\Bigl\{ 8
\s^2(1+\s^2)u^{10}+2(1+\s^2)(1-15\s^2+8\s^4)u^8 \nonumber\\
&-& (3-41\s^2-36\s^4)u^6 +
3\{2(1+\s^2)u_0^2[(1+\s^2)u_0^2-1]-7\s^2\}u^4 \nonumber\\
&-& \{(9+8\s^2)[(1+\s^2)u_0^2-1]-1\} u^2 + 3
u_0^2[(1+\s^2)u_0^2-1] \Bigr\}\ ,\nonumber\\
V_\psi(u;u_0,\s) &=& - {1 \ov 4u^2(u^2-1)[(1+\s^2)u^2-1]^3}
\Bigl\{ 16 \s^2(1+\s^2) u^{12} \\
&-& 2(1+\s^2)(1+35\s^2+18\s^4) u^{10} + (5+103\s^2+94\s^4+12\s^6)u^8 \nonumber\\
&+& \{6(1+\s^2)(1+2\s^2)u_0^2 [(1+\s^2)u_0^2-1]-3-54\s^2-20\s^4\}u^6 \nonumber\\
&-& \{(13+18\s^2+4\s^4)u_0^2[(1+\s^2)u_0^2-1]+1-7\s^2\}u^4 \nonumber\\
&+& \{8 u_0^2[(1+\s^2)u_0^2-1]+1\}u^2 - u_0^2[(1+\s^2)u_0^2-1]
\Bigr\}\ .\nonumber
\ea
\no $\bullet$ Sphere, $(\th_0,\psi_0)=(0,0 \hbox{ or } \pi/2)$:
\be
\label{B-13} V_\th(u;u_0,\s) = V_\th(u;u_0) + 4 \s^2 u^2\ ,\qq
V_\psi(u;u_0,\s) = V_\psi(u;u_0) + 4 \s^2 u^2\ ,
\ee
where $V_\th(u;u_0)$ and $V_\psi(u;u_0)$ are the potentials for
the corresponding undeformed case, with the former given by
\eqn{B-1} and the latter given by a function which, since the
$\psi$ fluctuations in the multicenter case are stable, admits no
negative-energy bound states. \no $\bullet$ Sphere,
$(\th_0,\psi_0)=(\pi/2,{\rm any})$:
\be
\label{B-14} V_\th(u;u_0,\s) = -1 + 4 \s^2(u^2-1)\ .
\ee
\no $\bullet$ Disc, $(\theta_0,\psi_0)=(0,\pi/4)$:
\ba
\label{B-15} V_\th(u;u_0,\s) &=& {1 \ov 4u^2[(1+\s^2)u^2+1]^3}
\Bigl\{ 8
\s^2(1+\s^2)u^{10} - 2(1+\s^2)(1-15\s^2-8\s^4)u^8 \nonumber\\
&-& (3-41\s^2-36\s^4)u^6 - 3\{2(1+\s^2)u_0^2[(1+\s^2)u_0^2+1]-7\s^2\}u^4 \nonumber\\
&-& \{(9+8\s^2)[(1+\s^2)u_0^2+1]-1\} u^2 - 3
u_0^2[(1+\s^2)u_0^2-1] \Bigr\}\ ,\nonumber\\
V_\psi(u;u_0,\s) &=& - {1 \ov 4u^2(u^2+1)[(1+\s^2)u^2+1]^3}
\Bigl\{16 \s^2(1+\s^2) u^{12} \\
&+& 2(1+\s^2)(1+35\s^2+18\s^4) u^{10} + (5+103\s^2+94\s^4+12\s^6)u^8 \nonumber\\
&-& \{6(1+\s^2)(1+2\s^2)u_0^2 [(1+\s^2)u_0^2+1]-3-54\s^2-20\s^4\}u^6 \nonumber\\
&-& \{(13+18\s^2+4\s^4)u_0^2[(1+\s^2)u_0^2+1]+1-7\s^2\}u^4 \nonumber\\
&-& \{8 u_0^2[(1+\s^2)u_0^2+1]+1\}u^2 - u_0^2[(1+\s^2)u_0^2+1]
\Bigr\}\ .\nonumber
\ea
\no $\bullet$ Disc, $(\theta_0,\psi_0)=(0,0 \hbox{ or } \pi/2)$:
\be
\label{B-16} V_\th(u;u_0,\s) = V_\th(u;u_0) + 4 \s^2 u^2\ ,\qq
V_\psi(u;u_0,\s) = V_\psi(u;u_0) + 4 \s^2 u^2\ ,
\ee
where $V_\th(u;u_0)$ and $V_\psi(u;u_0)$ are the potentials for
the undeformed case, with the former given by \eqn{B-3} and the
latter by a function admitting no negative-energy bound states.

\no $\bullet$ Disc, $(\th_0,\psi_0)=(\pi/2,{\rm any})$:
\be
\label{B-17} V_\th(u;u_0,\s) = 1 + 4 \s^2(u^2+1)\ .
\ee

\section{Angular fluctuations and the Lam\'e equation}

\no For the special cases of (i) the deformed sphere at $\th_0=0$
in the limit $\s \gg 1$ and (ii) the deformed sphere at
$\th_0=\pi/2$ (for any $\s$), the Schr\"odinger equation for the
angular fluctuations takes a relatively simple form that allows
for explicit solutions. In both cases, the Schr\"odinger variable
$x$ is given in closed form in terms of $u$ by Eq. \eqn{4-51},
which can be inverted for $u$ to yield
\be
u^2 = 2 u_0^2\left({1\ov {\rm sn}^2(\sqrt{2} u_0 x,k)}-\ha\right)
= 2 u_0^2 \left(\wp(\sqrt{2} u_0 x)+{1\ov 6 u_0^2}\right)\ ,
\label{asjh}
\ee
where ${\rm sn}(z,k)$ is the elliptic Jacobi function and $\wp(z)
\equiv \wp(z\,|\,\omega_1,\omega_2) \equiv \wp(z,g_2,g_3)$ is the
Weierstrass p-function, specified by the half-periods
\be
\label{A-4} \omega_1 = \elK(k)\ ,\qq \omega_2 = {\rm i} \elK(k')\
,\qq k^2={1+u_0^2\ov 2u_0^2}\ ,
\ee
or, equivalently, by the elliptic invariants
\be
\label{A-5} g_2 = -4 (e_1 e_2 + e_2 e_3 + e_3 e_1)\ ,\qq g_3 = 4
e_1 e_2 e_3\ ,
\ee
with the $e_i$ given by
\be
\label{A-6} e_1 = {3 u_0^2 - 1 \ov 6 u_0^2}\ ,\qq e_2 = {1 \ov 3
u_0^2}\ ,\qq e_3 = - {3 u_0^2 + 1 \ov 6 u_0^2}\ ,
\ee
and satisfying $e_1 + e_2 + e_3=0$. Substituting \eqn{asjh} into
\eqn{4-42} and \eqn{4-50} to obtain explicit expressions for the
potentials in terms of $x$, and making a further change of
variables to $z = \sqrt{2} u_0 x$, we arrive at the following
equation
\be
\label{A-1} \left\{ -{d^2 \ov dz^2} + \left[ A \wp(z) + h \right]
\right\} \Psi(z) = {\omega^2 \ov 2u_0^2} \Psi(z) \ ,\qq 0
\leqslant z \leqslant \omega_1\ ,
\ee
subject to the boundary conditions
\be
\label{A-2} \Psi(0) = 0 \ ,\qq \Psi^\prime(\omega_1)= 0 \ ,
\ee
and to the requirement that $\Psi(z)$ be real.
Here, $A$ and $h$ stand for the constants
\ba
\label{A-3} \hbox{ (i)} \: &:& \: A=2\phantom{\s^2}\ ,\qq h=-{5\ov
 3u_0^2}\ ,
\nonumber\\
 \hbox{(ii)} \: &:& \: A=4\s^2\  ,\qq h=-{3+4\s^2 \ov
6u_0^2}\ .
\ea

\no In the special cases where $A=n(n+1)$ (with $n$ being a
positive integer) Eq. \eqn{A-1} is just the Lam\'e equation, whose
exact solutions are well-known (see e.g. \cite{ince}). Although
the construction of the specific solutions satisfying the boundary
conditions \eqn{A-2} is rather cumbersome, it is instructive to
consider the simplest possible case, $n=1$, as a consistency check
of our numerics. For this case, the zero-mode equation simplifies
to
\be
\label{A-7} \left\{ -{d^2 \ov dz^2} + \left[ 2 \wp(z) + h \right]
\right\} \Psi(z) = 0\ ,\qq 0 \leqslant z \leqslant \omega_1\ .
\ee
For the analysis that follows, it is convenient to define the
quantities
\be
\label{A-8} a \equiv \wp^{-1}(h)\ ,\qq \omega_3 \equiv
\omega_1+\omega_2\ .
\ee
When $a \ne \omega_i$, $i=1,2,3$, the equation \eqn{A-7} possesses the
linearly-independent solutions
\be
\label{A-9} \Psi_\pm(z) = \exp \left[ \mp z \upzeta(a) \right]
{\upsigma(z \pm a) \ov \upsigma(z)}\ ,
\ee
where $\upzeta(z) \equiv \upzeta(z\,|\,\omega_1,\omega_2)$ and
$\upsigma(z) \equiv \upsigma(z\,|\,\omega_1,\omega_2)$ are the
Weierstrass zeta and sigma functions.\footnote{Some properties of
these functions, used in the derivation of \eqn{A-11}, are
\ban
&\upzeta^\prime(z)=-\wp(z)\ ,\qq
\upsigma^\prime(z)=\upsigma(z)\upzeta(z)\ , \nonumber\\
&\upzeta(z+2\omega_{1,2})= \upzeta(z)+2\upzeta(\omega_{1,2})\ ,\qq
\upsigma(z+2\omega_{1,2})=-\exp[2 (z+\omega_{1,2})
\upzeta(\omega_{1,2})]\upsigma(z)\ . \ean}When $a=\omega_i$, the
linearly-independent solutions are instead
\be
\label{A-10} \Psi_{i,1} (z) = \exp \left[ -z \upzeta(\omega_i)
\right] {\upsigma(z + \omega_i) \ov \upsigma(z)}\ , \qq \Psi_{i,2}
(z) = \left[ \upzeta(z+\omega_i)+ z e_i \right] \Psi_{i,1} (z)\ .
\ee
For these two cases, the general solution reads $\Psi(z) = C_+
\Psi_+(z) + C_- \Psi_-(z)$ and $\Psi(z) = C_{i,1} \Psi_{i,1}(z) +
C_{i,2} \Psi_{i,2}(z)$ respectively. Imposing the condition
$\Psi(0)=0$, we obtain $C_+ = C_-$ and $C_{i,1} =
-\upzeta(\omega_i) C_{i,2}$ respectively, while using the reality
condition for $\Psi(z)$ we find that $a$ must be real. For the
case $a \ne \omega_i$, the condition $\Psi^\prime(\omega_1)=0$
leads to the equation
\be
\label{A-11} \left\{ \upsigma(\omega_1-a) \left[ \upzeta(a) -
\upzeta(\omega_1) + \upzeta(\omega_1-a) \right] \right\}
\left\{\exp \left[ 2 \left( a \upzeta(\omega_1) - \omega_1
\upzeta(a) \right) \right] -1 \right\} = 0\ ,
\ee
which is to be solved for $a$ in the square $\{ 0 \leqslant \real
a < 2 \omega_1\ , \, 0 \leqslant \imag a < 2 |\omega_2| \}$ in the
complex plane, due to the periodicity of the Weierstrass functions
under $a \to a + 2m \omega_1 + 2n \omega_2$. The trivial zeros of
\eqn{A-11} occur at the points $a=\omega_i$, $i=1,2,3$ where
either the first or the second factor in curly brackets vanishes.
However, for these points, we actually have to use the second set
of linearly-independent solutions \eqn{A-10}. Doing so, we find
that the values $a=\omega_{2,3}$ are not allowed due to the
reality condition, while for the value $a=\omega_1$ the boundary
condition $\Psi^\prime(\omega_1)=0$ leads to the equation
$3\upzeta(\omega_1)=e_1\omega_1$, which has no solution.
Nontrivial roots of Eq. \eqn{A-11} can only arise from the second
factor in curly brackets, i.e. they are given by the solutions of
the equation
\be
\label{A-12} a \upzeta(\omega_1) = \omega_1 \upzeta(a)\ ,
\ee
subject to the condition $a \ne \omega_1$. Using Eqs. \eqn{A-3},
\eqn{A-4} and \eqn{A-8}, we can express this as a transcendental
equation for $u_{0 \rm c}$, whose solutions for the two cases of
interest are
\ba
\label{A-13} \hbox{ (i)} \: &:& \: u_{0 \rm c}\simeq 1.09737\ ,
\qq L_{\rm c}=1.9312\ ,
\nonumber\\
 \hbox{(ii)} \: &:& \: u_{0 \rm c} \simeq 1.02676\ ,\qq L_{\rm
 c}=2.92397\ ,\quad \hbox{for $4\s^2=2$}\ .
\ea
Both results are in precise agreement with our numerical analysis.

\section{Soap films and strings in Rindler space}

An example of the calculations considered here and in
\cite{stability1} has been presented recently in \cite{berenstein}
and refers to open strings ending on branes in Rindler space and
their stability properties. As is well-known, Rindler space is (a
portion of) flat Minkowski space, expressed in coordinates adapted
to an observer at constant acceleration. Such an observer
experiences the Unruh effect, perceiving the inertial vacuum as a
state populated by a thermal distribution of particles at a
temperature $T=\k/2\pi$, where $\k$ is the surface gravity. It is
then natural to expect that an open string with fixed endpoints
propagating in Rindler space will have similar properties as one
propagating in a black-hole background and, in particular, will
exhibit a phase structure of the type shown in Fig.~1c. In this
appendix we prove the amusing fact that this problem is {\em
exactly} equivalent to the classical-mechanical problem of the
shape of a soap film suspended between two circular rings.

\no The metric for Rindler space has the form
\be
\label{C-1} ds^2 = -\k^2 u^2 d t^2 + dy^2 + du^2 + \ldots\ ,
\ee
where $u$ is the radial direction with the Rindler horizon
corresponding to $u=0$ and $y$ is a generic spatial direction.
The string configuration of interest corresponds to an open string
with its two endpoints located at the same radius $u=\L$ and
separated by a distance $L$ along the $y$ direction. It is obvious
that the formalism of section 2 readily applies to this setup.
Passing to convenient dimensionless units through the rescalings
\be
\label{C-2} u \to \L u\ ,\qq u_0 \to \L u_0\ ,\qq L \to \L L\ ,\qq
E \to {\k \L^2\ov 4\pi^2} E\ ,
\ee
inserting the metric components of \eqn{C-1} in \eqn{2-10}, and
integrating from $u_0$ to $\L$, we find that the classical
solution is the catenary curve of Leibniz, Huygens and Bernoulli,
\be
\label{C-2a} u = u_0 \cosh {y \ov u_0}\ ,
\ee
which is a slice of Euler's catenoid. The integration constant
$u_0$ and the energy of the string are determined by Eqs.
\eqn{2-12} and \eqn{2-13} (without the subtraction term) which for
our case read
\be
\label{C-3} L = 2 u_0 \int_{u_0}^{1} {du \ov \sqrt{u^2-u_0^2}} = 2
u_0 \cosh^{-1} {1 \ov u_0}\ ,
\ee
and
\be
\label{C-4} E = 4 \pi \int_{u_0}^1 {d u \, u^2 \ov
\sqrt{u^2-u_0^2}} = 2 \pi \left( \sqrt{1-u_0^2} + u_0^2 \cosh^{-1}
{1 \ov u_0} \right)\ ,
\ee
respectively. Eqs. \eqn{C-2a}--\eqn{C-4} are exactly the same
formulas appearing in Plateau's problem for a thin soap film of
mass density $1$ and surface tension $\ha$ stretched between two
coaxial rings of radius $1$ that are separated by a distance $L$,
with $u_0$ being the minimal radius reached by the film (see Eqs.
(A.4)--(A.5) in appendix A of \cite{stability1}), and thus the
properties of the solutions (see e.g. \cite{variations}) are the
same. Namely, there exists a critical value $u_{0 \rm c}$ of $u_0$
and a corresponding value $L_{\rm c}$ of $L$ at which
$L^{\prime}(u_0)=0$, i.e.
\be
\label{C-5} \sqrt{1-u_0^2} \cosh^{-1} {1 \ov u_0} = 1\ .
\ee
Solving this equation, we find $u_{0 {\rm c}} \simeq 0.552$,
whence $L_{\rm c} \simeq 1.325$ (to compare with the results of
\cite{berenstein}, set $u_{0 \rm} \to L / h$). For separations
$L<L_{\rm c}$, there is a ``short'' string (dual to the shallow
catenoid), a ``long'' string of higher energy (dual to the deep
catenoid), and a configuration of two straight strings (dual to
the Goldschmidt solution). For separations $L>L_{\rm c}$, only the
last configuration is possible.

\no Regarding stability, the only relevant fluctuations are the
longitudinal fluctuations along $y$. By the results of section 2,
instabilities occur for $u_0$ below the critical value where
$L^{\prime}(u_0)=0$ i.e. for $u_0 < u_{0\rm c}$. This can be
verified directly by noting that the longitudinal Sturm--Liouville
equation has the form
\be
\label{C-6} (u^2-u_0^2) \d y^{\prime\prime}(u)+ {u^2+2u_0^2 \ov u}
\d y^{\prime}(u) = -{\omega^2 \ov \k^2} \d y(u)\ .
\ee
The zero-mode solution satisfying the boundary condition $\d
y(1)=0$ is given by
\be
\label{C-7} \d y (u) \sim \cosh^{-1} {u \ov u_0} - {u \ov
\sqrt{u^2-u_0^2}} - \cosh^{-1} {1 \ov u_0} + {1 \ov
\sqrt{1-u_0^2}}\ ,
\ee
and imposing the boundary condition $\lim_{u\to u_0^+} \left[ \d
y(u) + 2 (u-u_0) \d y^\prime(u) \right] = 0$ leads indeed to Eq.
\eqn{C-5}. Alternatively, one may set $x=\cosh^{-1}{u\ov u_0}$ and
$\Psi= \tanh x \ \d y$ to transform \eqn{C-6} to a Schr\"odinger
equation with the P\"oschl--Teller type II potential $V=-{2 \ov
\cosh^2 x}$ and the energy ${\omega^2\ov \k^2}$, defined in the
interval $x \in [1,\cosh^{-1}{1\ov u_0}]$ and obeying Neumann and
Dirichlet boundary conditions at the left and right endpoint
respectively.\footnote{This is different than the corresponding
Schr\"odinger equation for the fluctuations of the thin soap film
problem (see \cite{stability1,durand}).} The solutions are the
associated Legendre functions $P_1^\nu(\tanh x)$ and
$Q_1^\nu(\tanh x)$, where $\nu = {{\rm i} \omega \ov \k}$, and
imposing the boundary conditions on the zero-mode solutions leads
again to Eq. \eqn{C-5}. In both formulations, it is very
straightforward to check that the lowest eigenvalue really changes
sign from positive to negative as $u_0$ crosses $u_{0 \rm c}$ from
the right. In conclusion, the long strings with $u_0<u_{0 {\rm
c}}$ are perturbatively unstable in the region where they exist.

\no The qualitative analogy of the soap-film problem with AdS/CFT
calculations of Wilson loops has already been considered in
previous work \cite{stability1,gross}. It is a pleasant surprise
that this problem actually has an exact analog in the context of
the Nambu-Goto string.


\begin{thebibliography}{99}

\renewcommand{\baselinestretch}{1}
\normalsize


\bibitem{adscft} J.M.~Maldacena,
Adv.\ Theor.\ Math.\ Phys.\ {\bf 2} (1998) 231, Int.\ J.\ Theor.\
Phys.\  {\bf 38} (1999) 1113, {\tt hep-th/9711200}.\hfill\break
S.S.~Gubser, I.R.~Klebanov and A.M.~Polyakov,
Phys.\ Lett.\ {\bf B428} (1998) 105, {\tt
hep-th/9802109}.\hfill\break
E.~Witten,
Adv.\ Theor.\ Math.\ Phys.\ {\bf 2} (1998) 253, {\tt
hep-th/9802150}
and
Adv.\ Theor.\ Math.\ Phys.\  {\bf 2} (1998) 505, {\tt
hep-th/9803131}.\hfill\break
O.~Aharony, S.S.~Gubser, J.M.~Maldacena, H.~Ooguri and Y.~Oz,
Phys.\ Rept.\ {\bf 323} (2000) 183, {\tt hep-th/9905111}.


\bibitem{maldaloop} J.M.~Maldacena,
Phys. Rev. Lett. {\bf 80} (1998) 4859, {\tt
hep-th/9803002}.\hfill\break
S.J.~Rey and J.T.~Yee,
Eur. Phys. J. {\bf C22} (2001) 379, {\tt hep-th/9803001}.

\bibitem{wilsonloopTemp}
S.J.~Rey, S.~Theisen and J.T.~Yee,
Nucl.\ Phys.\ {\bf B527} (1998) 171, {\tt hep-th/9803135}.
\hfill\break
A.~Brandhuber, N.~Itzhaki, J.~Sonnenschein and S.~Yankielowicz,
Phys.\ Lett.\ {\bf B434} (1998) 36, {\tt hep-th/9803137}
and
JHEP {\bf 9806} (1998) 001, {\tt hep-th/9803263}.

\bibitem{bs}
A.~Brandhuber and K.~Sfetsos,
Adv. Theor. Math. Phys. {\bf 3} (1999) 851,\hfill\break {\tt
hep-th/9906201}.

\bibitem{sonnenschein}
Y.~Kinar, E.~Schreiber and J.~Sonnenschein,
Nucl.\ Phys.\ {\bf B566} (2000) 103, {\tt
hep-th/9811192}.\hfill\break
Y.~Kinar, E.~Schreiber, J.~Sonnenschein and N.~Weiss,
Nucl.\ Phys.\ {\bf B583} (2000) 76,  {\tt hep-th/9911123}.

\bibitem{sonnenschein2}
J.~Sonnenschein, {\em What does the string / gauge correspondence
teach us about Wilson loops?}, {\tt hep-th/0003032}.


\bibitem{cg}
C.G.~Callan and A.~Guijosa,
Nucl.\ Phys.\ {\bf B565} (2000) 157, {\tt hep-th/9906153}.

\bibitem{kmt}
I.R.~Klebanov, J.M.~Maldacena and C.B.~Thorn,
JHEP {\bf 0604} (2006) 024,\hfill\break {\tt hep-th/0602255}.


\bibitem{trivedi} P.~Kraus, F.~Larsen and S.P.~Trivedi,
JHEP {\bf 9903} (1999) 003, {\tt hep-th/9811120}.

\bibitem{sfet1} K.~Sfetsos,
JHEP {\bf 9901} (1999) 015, {\tt hep-th/9811167}.

\bibitem{warn} D.Z.~Freedman, S.S.~Gubser, K.~Pilch and N.P.~Warner,
JHEP {\bf 0007} (2000) 038, {\tt hep-th/9906194}.

\bibitem{Basfe2} I.~Bakas and K.~Sfetsos,
Nucl. Phys. {\bf B573} (2000) 768, {\tt
hep-th/9909041}.\hfill\break
I.~Bakas, A.~Brandhuber and K.~Sfetsos,
Adv. Theor. Math. Phys. {\bf 3} (1999) 1657, {\tt
hep-th/9912132}.\hfill\break
M.~Cvetic, S.S.~Gubser, H.~L\"u and C.N.~Pope,
Phys.\ Rev. {\bf D62} (2000) 086003, {\tt hep-th/9909121}.


\bibitem{stability1}
S.D.~Avramis, K.~Sfetsos and K.~Siampos,
Nucl.\ Phys.\  {\bf B769} (2007) 44,\hfill\break {\tt hep-th/0612139}.


\bibitem{psz}
K.~Peeters, J.~Sonnenschein and M.~Zamaklar,
Phys.\ Rev.\ {\bf D74} (2006) 106008, {\tt hep-th/0606195}.

\bibitem{lrw}
H.~Liu, K.~Rajagopal and U.A.~Wiedemann, {\it An AdS/CFT
calculation of screening in a hot wind}, {\tt hep-ph/0607062} and
JHEP {\bf 0703} (2007) 066, {\tt hep-ph/0612168}.

\bibitem{cgg}
M.~Chernicoff, J.A.~Garcia and A.~Guijosa,
JHEP {\bf 0609} (2006) 068,\hfill\break {\tt hep-th/0607089}.

\bibitem{aev}
P.C.~Argyres, M.~Edalati and J.F.~V\'azquez-Poritz,
JHEP {\bf 0701} (2007) 105, {\tt hep-th/0608118}.

\bibitem{cno}
E.~C\'aceres, M.~Natsuume and T.~Okamura,
JHEP {\bf 0610} (2006) 011,\hfill\break {\tt hep-th/0607233}.

\bibitem{asz1}
S.D.~Avramis, K.~Sfetsos and D.~Zoakos,
Phys.\ Rev.\  {\bf D75} (2007) 025009,\hfill\break {\tt
hep-th/0609079}.

\bibitem{naok}
M.~Natsuume and T.~Okamura, {\em Screening length and the
direction of plasma winds}, {\tt 0706.0086 [hep-th]}.


\bibitem{michalogiorgakis}
J.J.~Friess, S.S.~Gubser, G.~Michalogiorgakis and S.S.~Pufu,
JHEP {\bf 0704} (2007) 079, {\tt hep-th/0609137}.


\bibitem{cy}
M.~Cvetic and D.~Youm, Nucl. Phys. {\bf B477} (1996) 449, {\tt
hep-th/9605051}.

\bibitem{rs}
J.G.~Russo and K.~Sfetsos,
Adv.\ Theor.\ Math.\ Phys.\  {\bf 3} (1999) 131, {\tt
hep-th/9901056}.

\bibitem{spinningbranesthermo}
S.S.~Gubser,
Nucl. Phys. {\bf B551} (1999) 667, {\tt
hep-th/9810225}.\hfill\break
R.G.~Cai and K.S.~Soh,
Mod. Phys. Lett.  {\bf A14} (1999) 1895, {\tt
hep-th/9812121}.\hfill\break
M.~Cvetic and S.S.~Gubser,
JHEP {\bf 9907} (1999) 010, {\tt hep-th/9903132}.\hfill\break
T.~Harmark and N.A.~Obers,
JHEP {\bf 0001} (2000) 008, {\tt hep-th/9910036}.


\bibitem{LM} O.~Lunin and J.~Maldacena,
JHEP {\bf 0505} (2005) 033, {\tt hep-th/0502086}.

\bibitem{frolov}
S.A.~Frolov, R.~Roiban and A.A.~Tseytlin,
JHEP {\bf 0507} (2005) 045, {\tt hep-th/0503192}.\hfill\break
S.~Frolov,
JHEP {\bf 0505} (2005) 069, {\tt hep-th/0503201}.

\bibitem{marginal-generalizations}
R.C.~Rashkov, K.S.~Viswanathan and Y.~Yang,
Phys.\ Rev.\ {\bf D72} (2005) 106008, {\tt
hep-th/0509058}.\hfill\break
C.h.~Ahn and J.F.~V\'azquez-Poritz,
JHEP {\bf 0507} (2005) 032, {\tt hep-th/0505168}

\bibitem{LS} R.G.~Leigh and M.J.~Strassler,
Nucl. Phys. {\bf B447} (1995) 95, {\tt hep-th/9503121}.

\bibitem{hsz} R.~Hern\'andez, K.~Sfetsos and D.~Zoakos,
JHEP {\bf 0603}, 069 (2006), {\tt hep-th/0510132}
and
Fortsch.\ Phys.\ {\bf 54}, 407 (2006), {\tt hep-th/0512158}.

\bibitem{ahn-poritz}
C.~Ahn and J.F.~V\'azquez-Poritz,
JHEP {\bf 0606} (2006) 061, {\tt hep-th/0603142}.

\bibitem{asz2}
S.D.~Avramis, K.~Sfetsos and D.~Zoakos,
{\em Complex marginal
deformations of D3-brane geometries, their Penrose limits and
giant gravitons}, {\tt 0704.2067 [hep-th]}.


\bibitem{kt}
I.R.~Klebanov and A.A.~Tseytlin,
Nucl.\ Phys.\ {\bf B578} (2000) 123, {\tt hep-th/0002159}.

\bibitem{ks}
I.R.~Klebanov and M.J.~Strassler,
JHEP {\bf 0008} (2000) 052, {\tt hep-th/0007191}.

\bibitem{mn}
J.M.~Maldacena and C.~N\'u\~nez,
Phys.\ Rev.\ Lett.\  {\bf 86} (2001) 588, {\tt hep-th/0008001}.


\bibitem{AS1}
S.D.~Avramis and K.~Sfetsos,
JHEP {\bf 0701} (2007) 065, {\tt hep-th/0606190}.


\bibitem{LLQM}
L.D.~Landau and E.M.~Lifshitz, {\em Quantum Mechanics
(Non-relativistic theory)}, (Pergamon, London, 1959).

\bibitem{Callen}
H.B. Callen, {\em Thermodynamics and introduction to
thermostatistics}, 2nd edition, (John Wiley \& Sons, New York,
1985).

\bibitem{ince}
E.L.~Ince, {\em Ordinary Differential Equations}, Dover
Publications, New York, 1956.\hfill\break E.T.~Whittaker and
G.N.~Watson, {\em A Course of Modern Analysis}, Cambridge
University Press, Cambridge, 1986.

\bibitem{berenstein}
D.~Berenstein and H.J.~Chung,
{\em Aspects of open strings in
Rindler Space},\hfill\break {\tt 0705.3110 [hep-th]}.

\bibitem{variations}
C.~Caratheodory, {\em Calculus of Variations and Partial
Differential Equations of the First Order}, American Mathematical
Society, 1999.\hfill\break G.A.~Bliss, {\em Calculus of
Variations}, Chicago, IL: Open Court, 1925. \hfill\break
R.~Courant and H.~Robbins, {\em What is Mathematics?}, Oxford
University Press, Oxford, 1941. \hfill\break G.B.~Arfken and
H.J.~Weber, {\it Mathematical Methods for Physicists}, 6th
edition, Elsevier Academic Press.

\bibitem{gross}
D.J.~Gross and H.~Ooguri,
Phys.\ Rev.\ {\bf D58} (1998) 106002, {\tt hep-th/9805129}.

\bibitem{durand}
L.~Durand,
Am. J. Phys. {\bf 49} (1981) 334.


\end{thebibliography}
\end{document}